\documentstyle[amssymb,preprint,aps,epsf,psfig]{revtex}
\draft

\begin{document}
\title{Renormalization group analysis of magnetic and superconducting instabilities
near van Hove band fillings}
\author{A.A. Katanin$^{a,b}$ and A.P. Kampf$^a$}
\address{$^a$ Institut f\"ur Physik, Theoretische Physik III,\\
Elektronische Korrelationen und Magnetismus,\\
Universit\"at Augsburg, 86135 Augsburg, Germany\\
$^b$ Institute of Metal Physics, 620219 Ekaterinburg, Russia}
\maketitle

\begin{abstract}
Phase diagrams of the two-dimensional one-band $t$-$t^{\prime }$ Hubbard
model are obtained within the two-patch and the temperature-cutoff
many-patch renormalization group approach. At small $t^{\prime }$ and at van
Hove band fillings antiferromagnetism dominates, while with increasing $%
t^{\prime }$ or changing filling antiferromagnetism is replaced by $d$-wave
superconductivity. Near $t^{\prime }=t/2$ and close to van Hove band
fillings the system is unstable towards ferromagnetism. Away from van Hove
band fillings this ferromagnetic instability is replaced by a region with
dominating triplet $p$-wave superconducting correlations. The results of
the renormalization-group approach are compared with the mean-field results
and the results of the T-matrix approximation.
\end{abstract}


\vspace{0.2in}

\section{Introduction}

The close relation between antiferromagnetism (AF) and $d$-wave
superconductivity (dSC) was the subject of intensive investigations during
the last two decades (see e.g. Refs. \cite
{Scalapino,BSW,Zhang,Pines,Chubukov}). In particular, it was argued that the
superconducting properties of high-$T_c$ (HTSC) materials are intimately
related to their inherent antiferromagnetic correlations and many features
of these materials were explained from the point of view of competition
between antiferromagnetic and superconducting correlations \cite{Zhang}. On
the other hand, AF spin fluctuations also serve as the natural candidate for
the pairing mechanism of dSC\cite{Pines,Chubukov}. A distinctly different
physical situation is realized in the layered ruthenate Sr$_2$RuO$_4,$ which
is an unconvential and most likely triplet superconductor \cite{PTReview}.
It was proposed, that the pairing in this material results from
ferromagnetic spin fluctuations\cite{Mazin,Murakami}. Although inelastic
neutron scattering has so far been unsuccesful to detect significant
low-energy ferromagnetic spin fluctuations in this material\cite{Maeno},
this idea finds experimental support from the recent measurements of the
susceptibility of the electron doped compound Sr$_{2-x}$La$_x$RuO$_4$\cite
{ED} which revealed the tendency towards ferromagnetism with La doping.
Furthermore, the isoelectronic compound Ca$_2$RuO$_4$ also shows
ferromagnetism under hydrostatic pressure\cite{Maeno1}.

Both, copper-oxide systems and Sr$_2$RuO$_4$, are layered materials.\
Therefore both systems motivate the investigation of the competition and the
mutual interplay between magnetic and superconducting instabilities in
two-dimensional (2D) correlated electron systems. For this type of analysis
it is important to account for specific band structure related phenomena,
namely for the form of the Fermi surface (FS) and the electronic dispersion.
The influence of the shape of the FS on superconducting and magnetic
properties is of interest from both, theoretical and experimental points of
view and the theoretical analysis can be guided by material-specific
information obtained from angle-resolved photoemission (ARPES) experiments%
\cite{ARPES,ARPES1,ARPES2,Damascelli}.

The simplest theoretical model which allows to investigate the effect of the
band dispersion on magnetic ordering and superconductivity of 2D systems is
the single-band $t$-$t^{\prime }$ Hubbard model on a square lattice which
takes into account both nearest-neighbor $t$ and next-nearest-neighbor $%
t^{\prime }$ hopping. This model is often discussed in connection with HTSC
compounds, and it describes well the shape of the FSs of cuprate
superconductors observed in ARPES \cite{ARPES,ARPES1,ARPES2}. In particular,
the value $t^{\prime }/t=0.15$ was chosen for La$_2$CuO$_4$ and the value $%
t^{\prime }/t=0.30$ for the Bi2212 system \cite{t'/t} in the tight-binding
parametrization of the relevant electronic band for the CuO$_2$-planes,
although the realistic modelling of the latter bilayer material requires the
inclusion of interlayer hopping as well. On the other hand, Sr$_2$RuO$_4$
has three relevant bands\cite{SrBand}. Interband effects are not negligible
in this material, and may even prove important for the origin of
unconventional superconductivity \cite{ARS}.

Already in early mean-field and quantum Monte Carlo (QMC)\ studies of the $t$%
-$t^{\prime }$ Hubbard model \cite{LinHirsch} it was found that depending on
the ratio $t^{\prime }/t$ and the band filling, different types of
instabilities are possible. For small $t^{\prime }/t$ near half-filling the
FS is almost nested, which is the origin for antiferromagnetism in the
weak-coupling regime. $t^{\prime }$ hopping destroys the perfect nesting
property of the FS and therefore leads to ``frustration'' of
antiferromagnetism due to the hopping processes on the same sublattice and
may therefore favor the emergence of a superconducting state \cite{Santos}.
Furthermore, $t^{\prime }$ hopping also weakens the tendency towards stripe
formation\cite{Kampf} and by the suppression of this alternative instability
superconducting fluctuations may get enhanced. At the same time, larger
values of $t^{\prime }$ move the system closer to a ferromagnetic
instability, since for $t^{\prime }/t$ close to $1/2$ the dispersion is
flattened close to the bottom of the band. This leads to flat-band
ferromagnetism \cite{Tasaki} at low densities which was investigated earlier
for the $2D$ $t$-$t^{\prime }$ Hubbard model within the $T$-matrix
approximation \cite{Fleck,Hlubina} and projected QMC simulations \cite
{Hlubina1}.

The interplay of antiferromagnetism and $d$-wave superconductivity in the
one-band $t$-$t^{\prime }$ Hubbard model was recently reconsidered within
many-patch renormalization-group (RG) approaches\cite{Zanchi,Metzner,SalmHon}%
. It was indeed verified that with increasing $t^{\prime }$ and/or
decreasing band filling antiferromagnetism is replaced by $d$-wave
superconductivity. In the early RG approaches of Refs. \cite
{Zanchi,Metzner,SalmHon} particle-hole scattering at small momenta was not
treated on equal footing with other types of scattering, and therefore these
analyses were unable to search for a possible ferromagnetic instability (see
the discussion in Refs. \cite{VHIKK,SalmHon1}). It was shown in Ref. \cite
{Guinea} however, that particle-hole scattering at small momenta does indeed
lead to the appearance of a ferromagnetic phase at large enough $t^{\prime
}/t$ and at van Hove (vH) band fillings; the onset of dominant ferromagnetic
correlations was found to occur for $t^{\prime }/t>0.27$. However, unlike in
Refs.\cite{Zanchi,Metzner,SalmHon}, the contribution of the Cooper channel
was not taken into account in Ref. \cite{Guinea}. The possibility of a
ferromagnetic instability was also investigated within a simplified
two-patch RG scheme \cite{VHIKK}, which considers only the scattering of
electrons in the vicinity of the ``singular'' points $(\pi ,0)$ and $(0,\pi )
$ in momentum space and therefore gives only a rough picture for the RG
scaling behavior of the coupling constants. The temperature-cutoff version
of the many-patch RG approach (TCRG) recently introduced by Honerkamp and
Salmhofer \cite{SalmHon1} includes the contributions of the whole Brillouin
zone and uses the temperature as a natural low-energy cutoff parameter in
order to avoid the technical difficulties with the inclusion of
small-momentum particle-hole scattering. It was demonstrated in Refs. \cite
{VHIKK,SalmHon1} that the proper account of all scattering channels indeed
leads to ferromagnetism at large enough $t^{\prime }/t.$ Moreover, the
critical value of $(t^{\prime }/t)_c$ for the stability of ferromagnetism is
$U$-dependent \cite{VHIKK} (unlike the results of Ref. \cite{Guinea}), in
particular for $U\rightarrow 0$ ferromagnetism exists only in the flat-band
low-density limit $(t^{\prime }/t)_c\rightarrow 1/2$ in qualitative
agreement with the results of the $T$-matrix approximation for the effective
electron-electron interaction vertex\cite{Fleck}. Naturally, ferromagnetic
and $d$-wave superconducting fluctuations tend to suppress each other\cite
{VHIKK,SalmHon1}. The suppression of ferromagnetism by superconducting
fluctuations is reminiscent of the well-known Kanamori screening \cite
{Kanamori}. On the other hand, as shown in Ref. \cite{SalmHon1}, the
tendency towards {\it triplet} superconductivity is enhanced by
ferromagnetic fluctuations and may exist in the vicinity of a ferromagnetic
phase. Note that the Pomeranchuk instability, which was proposed for small $%
t^{\prime }$ in Ref. \cite{Metzner1}, was shown to be a non-leading
instability in the $t$-$t^{\prime }$ Hubbard model \cite{SalmHon02}.

With these recent results it appears as a natural task to investigate
systematically the weak-coupling phase diagram of the 2D Hubbard model
within the RG approach. Earlier, such an analysis was performed only with
the momentum cutoff RG versions \cite{Zanchi,Metzner,SalmHon}, which, as we
discussed above, do not allow to include the contribution of ferromagnetic
fluctuations. Previous studies of the model with the inclusion of all
electron scattering channels were performed either in the two-patch RG
scheme \cite{VHIKK}, which is restricted to vH band fillings or within the
TCRG approach\cite{SalmHon1,SalmHon02}, which however was applied only for
some selected parameter values and therefore results for a phase diagram in
the weak-coupling regime were not obtained.

The purpose of the present paper is to investigate systematically different
types of instabilities of the $t$-$t^{\prime }$ Hubbard model within
two-patch and the many-patch TCRG approach. The paper is organized as
follows. In Section II we give a summary of the RG methods we use. In
Section III we present the phase diagrams obtained and compare the results
with previous investigations of the $t-t^{\prime }$ Hubbard model. In
Section IV we discuss the results and conclude.

\section{Renormalization-group approaches}

We consider the $t$-$t^{\prime }$ Hubbard model
\begin{equation}
H=-\sum_{ij\sigma }t_{ij}c_{i\sigma }^{\dagger }c_{j\sigma
}+U\sum_in_{i\uparrow }n_{i\downarrow }-(\mu -4t^{\prime })N
\end{equation}
where the hopping amplitude $t_{ij}=t$ for nearest neighbor sites $i$ and $j$
and $t_{ij}=-t^{\prime }$ for next-nearest neighbor sites ($t,t^{\prime }>0$%
) on a square lattice (we have shifted the chemical potential $\mu $ by $%
4t^{\prime }$ for further convenience). In momentum space Eq. (1) reads
\begin{equation}
H=\sum_{{\bf k}\sigma }\varepsilon _{{\bf k}}c_{{\bf k}\sigma }^{\dagger }c_{%
{\bf k}\sigma }+\frac U{2N^2}\sum_{{\bf k}_1{\bf k}_2{\bf k}_3{\bf k}%
_4}\sum_{\sigma \sigma ^{\prime }}c_{{\bf k}_1\sigma }^{\dagger }c_{{\bf k}%
_2\sigma ^{\prime }}^{\dagger }c_{{\bf k}_3\sigma ^{\prime }}c_{{\bf k}%
_4\sigma }\delta _{{\bf k}_1+{\bf k}_2-{\bf k}_3-{\bf k}_4}  \label{H1}
\end{equation}
where the Kronecker $\delta $-symbol ensures momentum conservation and the
dispersion has the form
\begin{equation}
\varepsilon _{{\bf k}}=-2t(\cos k_x+\cos k_y)+4t^{\prime }(\cos k_x\cos
k_y+1)-\mu   \label{ek}
\end{equation}
where the lattice constant is set to unity. The tight-binding spectrum (\ref
{ek}) leads to vH singularities (vHS) in the density of states arising from
the contributions around the points ${\bf k}_A=(\pi ,0)\ $and ${\bf k}%
_B=(0,\pi ).$ These singularities lie at the FS if $\mu =0$. For $t^{\prime
}=0$ the corresponding filling is $n_{VH}=1$ and the FS is nested, but the
nesting is removed for any $t^{\prime }\neq 0.$ The dependence of the vH
band filling on $t^{\prime }$ is shown in Fig.1, and the shape of the FS at
different $t^{\prime }/t$ and vH band fillings is shown in Fig. 2.

The standard RG strategy for fermion systems\cite{Shankar} is to integrate
out step by step the electronic states which are far from the FS (i.e. the
states with the energy $\Lambda -d\Lambda <\varepsilon _{{\bf k}}<\Lambda $
at each RG step). This procedure meets a difficulty when it is applied to a
FS with singular points, i.e. the points ${\bf k}_F^s$ with vanishing Fermi
velocity $\nabla \varepsilon _{{\bf k}}|_{{\bf k}={\bf k}_F^s}={\bf 0},$ as
in points ${\bf k}_A$ and ${\bf k}_B.$ In this case, the states with the
same excitation energy $\varepsilon _{{\bf k}}$ become inequivalent: the
excitations with momenta closer to the singular points produce more
divergent contributions to the renormalization of the electron-electron
interaction vertices than the excitations with momenta far from the singular
points. Therefore, an additional separation of the momenta besides the
standard separation into ``slow'' ($\varepsilon _{{\bf k}}<\Lambda $) and
``fast'' ($\varepsilon _{{\bf k}}>\Lambda $) modes is needed. The two-patch
approach which we consider in Sect. IIIa accounts only for the most singular
contributions coming from the immediate vicinities of the singular points.
The more sophisticated many-patch approaches of Refs. \cite
{Zanchi,Metzner,SalmHon,SalmHon1} (see Sect. IIIb) take into account the
momentum dependence of the interaction in a more accurate way by introducing
a set of patches which cover the entire Brillouin zone and parametrize the
interactions by the position of incoming and outgoing momenta on the patched
FS.

\subsection{Two-patch renormalization-group approach}

The two-patch approach\cite{Led,FurRice,VHIKK} is restricted to the vH band
fillings only. At these fillings the density of states at the Fermi energy
and the electron-electron interaction vertices at momenta ${\bf k}_i={\bf k}%
_{A,B}$ contain logarithmic divergencies arising from the momentum
integrations in the vicinity of the points ${\bf k}_{A,B}$. Therefore these
contributions are the most important for the calculation of the renormalized
interaction vertices. Accordingly, we subdivide the momentum space into
three types of regions (see Fig. 3). Regions I with ${\bf k}\in O(A)\vee O(B)
$ where
\begin{equation}
O(A)=\{{\bf k}:\,|{\bf k}-{\bf k}_A|<\Lambda \wedge |\varepsilon _{{\bf k}%
}/t|>e^{-\Lambda /|{\bf k}-{\bf k}_A|}\}
\end{equation}
and similar for $O(B)$ ($\Lambda $ is a momentum cutoff parameter) produce
the most singular contribution to the renormalization of the vertices.
Regions II contain the electronic states which are close to the FS but far
from vH singularities. It can be proven that the contributions of regions II
to the renormalization of the vertices is subleading in comparison with the
contributions of regions I, provided that $t^{\prime }/t$ is not small, i.e.
if the nesting effects are not important. Finally, regions III contain the
excitations which are far from both, the FS and vHS and do not produce
diverging contributions to any quantity. Therefore, in the simplest
approximation it is reasonable to neglect the contributions of regions II
and III altogether. A more accurate treatment within the many-patch RG
approach will be performed in the next section.

To account for the excitations with momenta in regions I, it is convenient
to introduce new electron operators $a_{{\bf k}}$ and $b_{{\bf k}}$ by
\[
c_{{\bf k}\sigma }=\left\{
\begin{array}{cc}
a_{{\bf k-k}_A,\sigma } & {\bf k}\in O(A) \\
b_{{\bf k-k}_B,\sigma } & {\bf k}\in O(B)
\end{array}
\right. .
\]
For momenta ${\bf k}\in O(A)\vee O(B)$ in the vicinity of the vH points the
dispersion is expanded as
\begin{mathletters}
\begin{eqnarray}
\varepsilon _{{\bf k}_A+{\bf p}} &\equiv &\varepsilon _{{\bf p}}^A=-2t(\sin
^2\varphi \,p_x^2-\cos ^2\varphi \,p_y^2)-\mu   \label{eka} \\
\  &=&-2tp_{+}p_{-}-\mu   \nonumber \\
\varepsilon _{{\bf k}_B+{\bf p}} &\equiv &\varepsilon _{{\bf p}}^B=2t(\cos
^2\varphi \,p_x^2-\sin ^2\varphi \,p_y^2)-\mu   \label{ekb} \\
\  &=&2t\widetilde{p}_{+}\widetilde{p}_{-}-\mu   \nonumber
\end{eqnarray}
where $\cos (2\varphi )=R=2t^{\prime }/t,\ p_{\pm }=p_x\sin \varphi \pm
p_y\cos \varphi ,$ and $\widetilde{p}_{\pm }=p_x\cos \varphi \pm p_y\sin
\varphi .$ Using the new electron operators we write the effective
Hamiltonian in the form
\end{mathletters}
\begin{eqnarray}
H &=&\sum_{{\bf p}\sigma }\varepsilon _{{\bf p}}^Aa_{{\bf p}\sigma
}^{\dagger }a_{{\bf p}\sigma }+\sum_{{\bf p}\sigma }\varepsilon _{{\bf p}%
}^Bb_{{\bf p}\sigma }^{\dagger }b_{{\bf p}\sigma }  \nonumber \\
&&+\frac{2\pi ^2t}{N^2}\sum_{{\bf p}_i,\sigma \sigma ^{\prime }}[g_1(\lambda
)a_{{\bf p}_1\sigma }^{\dagger }b_{{\bf p}_2\sigma ^{\prime }}^{\dagger }a_{%
{\bf p}_3\sigma ^{\prime }}b_{{\bf p}_4\sigma }+g_2(\lambda )a_{{\bf p}%
_1\sigma }^{\dagger }b_{{\bf p}_2\sigma ^{\prime }}^{\dagger }b_{{\bf p}%
_3\sigma ^{\prime }}a_{{\bf p}_4\sigma }]\delta _{{\bf p}_1+{\bf p}_2-{\bf p}%
_3-{\bf p}_4}  \nonumber \\
&&+\frac{\pi ^2t}{N^2}\sum_{{\bf p}_i,\sigma \sigma ^{\prime }}[g_3(\lambda
)a_{{\bf p}_1\sigma }^{\dagger }a_{{\bf p}_2\sigma ^{\prime }}^{\dagger }b_{%
{\bf p}_3\sigma ^{\prime }}b_{{\bf p}_4\sigma }+g_4(\lambda )a_{{\bf p}%
_1\sigma }^{\dagger }a_{{\bf p}_2\sigma ^{\prime }}^{\dagger }a_{{\bf p}%
_3\sigma ^{\prime }}a_{{\bf p}_4\sigma }+a
\begin{array}{c}
\leftrightarrow
\end{array}
b]\delta _{{\bf p}_1+{\bf p}_2-{\bf p}_3-{\bf p}_4}  \label{HVH}
\end{eqnarray}
where
\begin{equation}
\lambda =\ln (\Lambda /\max (p_{i+},p_{i-},\widetilde{p}_{i+},\widetilde{p}%
_{i-},T/t);  \label{al}
\end{equation}
the summation in Eq. (\ref{HVH}) is restricted to momenta ${\bf p}_i$ with $|%
{\bf p}_i|<\Lambda \ $and $|\varepsilon _{{\bf k}_{A,B}+{\bf p}%
_i}/t|>e^{-\Lambda /|{\bf p}_i|}.$

As shown in Fig. 4, the vertices $g_1$ to $g_4$ represent different types of
scattering processes of electrons with momenta close to the vHS. The bare
value for all four vertices is $g_i^0=U/(4\pi ^2t).$ The momentum dependence
of the vertex inside regions I is accounted for through the scaling variable
$\lambda $ only. Note however, that the momentum dependence of the
electronic spectrum within each patch is correctly taken into account in the
two-patch approach.

To obtain the dependence of the vertices $g_i$ on $\lambda $ we integrate
out at each RG step the fermions $a_{{\bf p}}$ with momenta $\Lambda
e^{-\lambda }<p_{\pm }<\Lambda e^{-\lambda -d\lambda },$ and fermions $b_{%
{\bf p}}$ with momenta $\Lambda e^{-\lambda }<\widetilde{p}_{\pm }<\Lambda
e^{-\lambda -d\lambda }$ (see the detailed description in Ref. \cite{UVJ}).
As argued above we neglect the renormalization of the $g_i$ arising from the
regions II and III of Fig. 3 since this leads to subleading corrections at
weak coupling. We determine the RG equations for the vertices $g_i(\lambda )$
in the form \cite{Led,FurRice,UVJ,VHIKK}
\begin{eqnarray}
{\rm d}g_1/{\rm d}\lambda &=&2d_1(\lambda
)g_1(g_2-g_1)+2d_2g_1g_4-2\,d_3g_1g_2,  \nonumber \\
{\rm d}g_2/{\rm d}\lambda &=&d_1(\lambda
)(g_2^2+g_3^2)+2d_2(g_1-g_2)g_4-d_3(g_1^2+g_2^2),  \nonumber \\
{\rm d}g_3/{\rm d}\lambda &=&-2d_0(\lambda )g_3g_4+2d_1(\lambda
)g_3(2g_2-g_1),  \nonumber \\
{\rm d}g_4/{\rm d}\lambda &=&-d_0(\lambda
)(g_3^2+g_4^2)+d_2(g_1^2+2g_1g_2-2g_2^2+g_4^2),  \label{TwoPatch}
\end{eqnarray}
where
\begin{eqnarray}
d_0(\lambda ) &=&2\lambda /\sqrt{1-R^2},\;d_2=2/\sqrt{1-R^2};\;  \nonumber \\
d_3 &=&2\tan ^{-1}(R/\sqrt{1-R^2})/R;  \nonumber \\
d_1(\lambda ) &=&2\min \{\lambda ,\ln [(1+\sqrt{1-R^2})/R]\}.
\end{eqnarray}
Eqs. (\ref{TwoPatch}) have to be solved with the initial conditions $%
g_i(0)=g_i^0$.

The authors of Ref. \cite{Guinea} argued that the kinematic restrictions
lead to the absence of particle-particle scattering contributions to the
vertices $g_i$ ($d_0=d_3=0$ in our notations). This conclusion however is
connected with the difficulty of the infinitesimal version of Wilson's RG
approach with a sharp momentum cutoff, since this approach does not allow to
treat correctly the vertices with nonzero momentum transfer and gives
artificially no renormalization for such vertices\cite{Wilson}. That is why
in the present approach we consider the vertices at the special vH points
only, rather than considering vertices with arbitrary momentum transfers.
Note that this difficulty does not arise in Wilson's RG approach with a
smooth cutoff and/or discrete RG transformations \cite{Wilson}; in this case
particle-particle scattering does contribute to the renormalization of the
vertices with arbitrary momenta.

In order to explore the possible instabilities of the system, we consider
the behavior of the zero-frequency, time-ordered response functions
\begin{equation}
\chi _m=\int\limits_0^{1/T}{\rm d}\tau \langle {\cal T\,}[\widehat{O}%
_m^{\dagger }(\tau )\widehat{O}_m(0)]\rangle
\end{equation}
in the zero-temperature limit $T\rightarrow 0.$ $\widehat{O}_m(\tau )$ are
the following operators
\begin{eqnarray}
\widehat{O}_{{\rm AF}} &=&\frac 1N\sum_{{\bf k}}\sigma c_{{\bf k},\sigma
}^{\dagger }c_{{\bf k+Q},\sigma },  \nonumber \\
\widehat{O}_{{\rm dSC}} &=&\frac 1N\sum_{{\bf k}\sigma }f_{{\bf k}}\sigma c_{%
{\bf k,}\sigma }^{\dagger }c_{-{\bf k},-\sigma }^{\dagger },  \nonumber \\
\widehat{O}_{{\rm F}} &=&\frac 1N\sum_{{\bf k}\sigma }\sigma c_{{\bf k,}%
\sigma }^{\dagger }c_{{\bf k,}\sigma },  \label{O}
\end{eqnarray}
in the Heisenberg representation, ${\cal T}$ is the imaginary time ordering
operation, and ${\bf Q}=(\pi ,\pi ).$ The order parameters which correspond
to $p$-wave pairing
\begin{equation}
\widehat{O}_{{\rm pSC}}^{x,y}=\frac 1N\sum_{{\bf k}\sigma }h_{{\bf k}%
}^{x,y}c_{{\bf k,}\sigma }^{\dagger }c_{-{\bf k},-\sigma }^{\dagger },
\end{equation}
with $h_{{\bf k}}^{x,y}=\sin k_{x,y}$ are irrelevant with the restriction of
momenta to the vicinities of vH points, since $h_{{\bf k}_{A,B}}=0$ and
therefore the possibility of triplet pairing can not be considered within
the two-patch approach. However, these order parameters can be taken into
account in many-patch approaches (see Section IIIb).

Picking up the logarithmical divergences in $\lambda $ we obtain the RG
equations for the dimensionless susceptibilities $\overline{\chi }_m=2\pi
^2t\chi _m$ in the same approximations as discussed above (cf. Refs. \cite
{Led,FurRice,VHIKK}):
\begin{eqnarray}
{\rm d}\overline{\chi }_m(\lambda )/{\rm d}\lambda  &=&d_{a_m}(\lambda )%
{\cal R}_m^2(\lambda ),  \label{hi} \\
{\rm d}\ln {\cal R}_m(\lambda )/{\rm d}\lambda  &=&d_{a_m}(\lambda )\Gamma
_m(\lambda ),  \nonumber
\end{eqnarray}
where the coefficients $\Gamma _m$ ($m=\,$AF, dSC, or F) are given by
\begin{equation}
\begin{array}{ll}
\Gamma _{{\rm AF}}=g_2+g_3; & \Gamma _{{\rm F}}=g_1+g_4; \\
\Gamma _{{\rm dSC}}=g_3-g_4. &
\end{array}
\label{GG}
\end{equation}
In Eqs. (\ref{hi}) $a_{\text{dSC}}=0$, $a_{\text{AF}}=1$, and $a_{\text{F}}=2
$. Eqs. (\ref{hi}) have to be solved with the initial conditions ${\cal R}%
_m(0)=1,$ $\chi _m(0)=0.$

The numerical solutions of Eqs. (\ref{TwoPatch}) show, that at a critical
value $\lambda _c$ of the scaling parameter $\lambda $ some of the vertices
and susceptibilities are divergent$.$ For a given $\lambda _c$ the size $%
\Lambda $ of the patches is restricted by $\ln (4/\Lambda )\ll \lambda _c.$
The latter criterion follows from the condition that the contribution of the
electrons with $|k_{\pm }|<\Lambda $ to particle-hole and particle-particle
bubbles is dominant (see e.g. Ref. \cite{Binz}). We choose $\Lambda =1$ and
require $\lambda _c\gg \ln 4\simeq 1.$ Since $\lambda _c$ decreases with
increasing interaction strength, this criterion defines the interaction
range where the two-patch RG approach is valid.

As an example, we show in Figs. \ref{g1},\ref{g2} the result of the
numerical solutions of Eqs. (\ref{TwoPatch}) for $U=2t$ and two different
choises of $t^{\prime }/t=0.1$ and $t^{\prime }/t=0.45$. The behavior of the
coupling constants is qualitatively different in these two cases. While in
the first case we have $g_{2,3}$ flowing to $+\infty $, $g_4$ to $-\infty ,$
and $g_1$ is mostly unchanged during the RG flow (we denote the
corresponding combination ($m++-$), the signs correspond to the behavior of
the coupling constants $g_1$-$g_4,$ $m$ means marginal. In the second case $%
g_{1,2,4}$ grow to $+\infty $ while $g_3$ goes to zero, i.e. we observe ($%
++0+$) behavior of the coupling constants. The comparison of the
corresponding susceptibilities shows that for $t^{\prime }/t=0.1$ the
antiferromagnetic susceptibility is the most divergent, while in the case $%
t^{\prime }/t=0.45$ the ferromagnetic susceptibility dominates and therefore
the two different coupling constant flows reflect two different
instabilities of the system. We discuss the complete phase diagram at vH
band fillings in Sect. IIIa.

\subsection{Many-patch renormalization-group analysis}

In the many-patch analysis we follow the temperature-cutoff RG for
one-particle irreducible Green functions proposed recently by Honerkamp and
Salmhofer in Ref. \cite{SalmHon1}. This version of the RG uses the
temperature as a natural cutoff parameter, allowing to account for both the
excitations with momenta close to the FS and far from it, which is necessary
for the description of instabilities which arise from zero-momentum
particle-hole scattering, e.g. ferromagnetism. Neglecting the frequency
dependence of the vertices, which is considered not important in the
weak-coupling regime, the RG differential equation for the temperature- and
momentum-dependent electron-electron interaction vertex has the form\cite
{SalmHon1} (see Fig. \ref{Diagram})
\begin{eqnarray}
&&\ \ \ \frac{{\rm d}}{{\rm d}T}V_T({\bf k}_1,{\bf k}_2,{\bf k}_3)
\begin{array}{c}
=
\end{array}
-\frac 1N\sum_{{\bf p}}V_T({\bf k}_1,{\bf k}_2,{\bf p})L_{\text{pp}}({\bf p}%
,-{\bf p}+{\bf k}_1+{\bf k}_2)V_T({\bf p},-{\bf p}+{\bf k}_1+{\bf k}_2,{\bf k%
}_3)  \nonumber \\
&&\ \ \;\,\,\,\,\,\,\,\,\,+\frac 1N\sum_{{\bf p}}\big[ -2V_T({\bf k}_1,{\bf p%
},{\bf k}_3)V_T({\bf p}+{\bf k}_1-{\bf k}_3,{\bf k}_2,{\bf p})+V_T({\bf k}_1,%
{\bf p},{\bf k}_3)V_T({\bf k}_2,{\bf p}+{\bf k}_1-{\bf k}_3,{\bf p})
\nonumber \\
&&\ \ \ \ \ \ \ \ \ +V_T({\bf k}_1,{\bf p},{\bf p}+{\bf k}_1-{\bf k}_3)V_T(%
{\bf p}+{\bf k}_1-{\bf k}_3,{\bf k}_2,{\bf p})\big] L_{\text{ph}}({\bf p},%
{\bf p}+{\bf k}_1-{\bf k}_3)  \nonumber \\
&&\ \ \ \ \ \ \ \ \ +\sum_{{\bf p}}V_T({\bf k}_1,{\bf p}+{\bf k}_2-{\bf k}_3,%
{\bf p})L_{\text{ph}}({\bf p},{\bf p}+{\bf k}_2-{\bf k}_3)V_T({\bf p},{\bf k}%
_2,{\bf k}_3)  \label{dV}
\end{eqnarray}
where
\begin{eqnarray}
L_{\text{ph}}({\bf k},{\bf k}^{\prime }) &=&\frac{f_T^{\prime }(\varepsilon
_{{\bf k}})-f_T^{\prime }(\varepsilon _{{\bf k}^{\prime }})}{\varepsilon _{%
{\bf k}}-\varepsilon _{{\bf k}^{\prime }}},  \nonumber \\
L_{\text{pp}}({\bf k},{\bf k}^{\prime }) &=&\frac{f_T^{\prime }(\varepsilon
_{{\bf k}})+f_T^{\prime }(\varepsilon _{{\bf k}^{\prime }})}{\varepsilon _{%
{\bf k}}+\varepsilon _{{\bf k}^{\prime }}},
\end{eqnarray}
and $f_T^{\prime }(\varepsilon )={\rm d}f(\varepsilon )/{\rm d}T,$ $%
f(\varepsilon )$ is the Fermi function. Eq. (\ref{dV}) has to be solved with
the initial condition $V_{T_0}({\bf k}_1,{\bf k}_2,{\bf k}_3)=U$ where the
initial temperature $T_0$ is of the order of the bandwidth$.$

The evolution of the vertices with decreasing temperature determines the
temperature dependence of the susceptibilities according to\cite
{SalmHon1,SalmHon02}
\begin{eqnarray}
\frac{{\rm d}}{{\rm d}T}\chi _{mT} &=&\sum_{{\bf k}^{\prime }}{\cal R}_{mT}(%
{\bf k}^{\prime }){\cal R}_{mT}(\mp {\bf k}^{\prime }+{\bf q}_m)L_{\text{pp},%
\text{ph}}({\bf k}^{\prime },\mp {\bf k}^{\prime }+{\bf q}_m),  \label{dH} \\
\frac{{\rm d}}{{\rm d}T}{\cal R}_{mT}({\bf k}) &=&\mp \sum_{{\bf k}^{\prime
}}{\cal R}_{mT}({\bf k}^{\prime })\Gamma _{mT}({\bf k},{\bf k}^{\prime })L_{%
\text{pp},\text{ph}}({\bf k}^{\prime },\mp {\bf k}^{\prime }+{\bf q}_m)
\nonumber
\end{eqnarray}
where
\begin{equation}
\Gamma _{mT}({\bf k,k}^{\prime })=\left\{
\begin{array}{cl}
V_T({\bf k},{\bf k}^{\prime },{\bf k}^{\prime }+{\bf q}_m) & \text{for }m=%
\text{AF or F} \\
V_T({\bf k},-{\bf k},{\bf k}^{\prime }) & \text{for }m=\text{dSC or pSC}
\end{array}
.\right.
\end{equation}
${\bf q}_m={\bf Q}$ for the AF susceptibility and ${\bf q}_m={\bf 0}$
otherwise. The upper signs and $pp$ indices in Eq. (\ref{dH}) refer to the
superconducting instabilities (dSC and pSC), and the lower signs and $ph$
indices to the other susceptibilities. The initial conditions for Eqs. (\ref
{dH}) are
\begin{equation}
{\cal R}_{m,T_0}({\bf k})=\left\{
\begin{array}{cl}
\cos k_x-\cos k_y & \text{for dSC} \\
\sin k_{x,y} & \text{for pSC} \\
1 & \text{otherwise}
\end{array}
,\right.   \label{T}
\end{equation}
and $\chi _{m,T_0}=0.$ To solve numerically Eqs. (\ref{dV}) and (\ref{dH}),
we use the discretization of momentum space in $N_p=48$ patches and the same
patching scheme as proposed in Ref. \cite{SalmHon1}. By exploiting the
symmetries of the square lattice, this reduces the above
integro-differential equations (\ref{dV}) and (\ref{dH}) to a set of 5824
differential equations which were solved numerically. We use the value of
the starting temperature $T_0=12t,$ which is slightly larger than the
bandwidth, and we stop the flow of the coupling constants at the temperature
$T_X$ when the maximum absolute value of the vertex function is larger than $%
V_{\max }=18t$. Note that the initial ${\bf k}$-dependence of the vertices (%
\ref{T}) is slightly changed during the RG flow: the vertices which
correspond to $d$- and $p$-wave superconductivity acquire $g$- and $f$-wave
(and even higher order) harmonics, respectively, and the vertices which have
$s$-wave symmetry (F, AF) acquire an additional extended $s$-wave ($\cos
k_x+\cos k_y$) component. However, these additional corrections are small.

To extract more detailed information about different instabilities we
extrapolate the inverse susceptibilities $\chi _{mT}^{-1}$ to temperatures
lower than $T_X$. For magnetically ordered or superconducting ground states
the corresponding extrapolated inverse susceptibilities $\chi _{mT}^{-1}$
vanish at some temperature $T_m^{*}$ below $T_X.$ The vanishing of the
inverse susceptibilities at finite temperatures is an artifact of the
one-loop RG approach and should be understood to determine a crossover
temperature into a renormalized classical regime with strong magnetic or
superconducting fluctuations and exponentially large correlation length $\xi
\propto \exp (A/T)$ for F and AF instabilities (see Appendix) and $\xi
\propto \exp (A/\sqrt{T-T_{\text{BKT}}})$ for superconducting instabilities.
$T_{\text{BKT}}$ is the Berezinskii-Kosterlitz-Thouless transition
temperature which arises because the superconducting transition in 2D is in
the same symmetry class as the classical 2D XY model \cite{BKT,Kogut}.
Although the inverse magnetic susceptibilities must be finite at $T<T_{\text{%
F,AF}}^{*},$ they are exponentially small, since in this regime $\chi _{%
\text{F,AF}}^{-1}\propto \xi ^{-2}$, cf. Refs. \cite{CHN,Kopietz,Vilk}. The
same concerns the superconducting susceptibilities in the temperature range $%
T_{\text{BKT}}<T<T_{\text{dSC,pSC}}^{*},$ where critical behavior $\chi _{%
\text{dSC,\thinspace pSC}}^{-1}\propto \xi ^{-2+\eta }$ ($\eta \cong 1/4$ is
the critical exponent for the susceptibility) as for the XY model\cite{BKT}
is expected. At $T<T_{\text{BKT}}$ the superconducting correlations have
power law decay in real space and the inverse static uniform order parameter
susceptibility does indeed vanish.

\section{Results and phase diagrams}

\subsection{van Hove band fillings}

First we trace the vH band fillings at different $t^{\prime }$, which are
determined by the condition $\mu =0$. The corresponding phase diagram in $%
t^{\prime }-U$ coordinates is plotted in Fig. \ref{vanHove}. Solid lines
correspond to the phase boundaries obtained within the two-patch approach,
while the symbols show different types of instabilities obtained within the
many-patch RG scheme\cite{Note}. Our many-patch results for $U=3t$ agree
quantatively with those obtained previously in Ref. \cite{SalmHon1}.

In both approaches we find antiferromagnetism for small $t^{\prime }$,
ferromagnetism for $t^{\prime }$ close to $1/2$ and $d$-wave
superconductivity for intermediate $t^{\prime }.$ The $t^{\prime }$ range
with a tendency towards $d$SC decreases with increasing $U.$ For
intermediate $t^{\prime }/t$ the susceptibilities with momenta ${\bf Q=}(\pi
-\delta ,\pi )$ are stronger than the antiferromagnetic susceptibility. The
incommensurate magnetic regions are indicated in Fig. \ref{vanHove} as well.
In the region $t^{\prime }/t\sim 0.3$ the behavior of the coupling constants
in the two-patch approach becomes ``frustrated'', namely all $g_i\rightarrow
0$ with increasing $\lambda .$ This frustration is the consequence of the
competition of antiferromagnetic and superconducting instabilities from one
side and the ferromagnetic instability from the other side and therefore at $%
t^{\prime }/t>0.2$ neither the antiferromagnetic nor the superconducting
susceptibility diverges in the two-patch approach$.$ The many-patch approach
suffers less from this problem; the frustrated behavior of the vertices is
observed only very close to the boundary of ferromagnetic and
antiferromagnetic or superconducting phases.

The boundary to the ferromagnetic phase appears almost identical in two- and
many-patch approaches, but the two-patch approach fails to reproduce the
location of the phase boundary between the antiferromagnetic and
superconducting phases for small $t^{\prime }$. This is similar to the
results for the extended $U$-$V$-$J$ Hubbard model\cite{UVJ}. As mentioned
in Section IIa, this difference is traced to the same behavior $(m++-)$ of
the coupling constants on approaching AF and dSC instabilities, while the
ferromagnetic phase is signalled by a different behavior of the coupling
constants $(++0+).$ Furthermore, the near-nesting effects, which are not
accounted for in the two-patch RG approach become particularly important at
small $t^{\prime }/t.$

We also mark in Fig. \ref{vanHove} the result of Alvarez et al. \cite{Guinea}
for the boundary of the ferromagnetic phase, obtained by neglecting the
contribution of particle-particle scattering. In this case the corresponding
two-patch RG equations can be solved analytically, since the coefficients $%
d_i$ in Eq. (\ref{TwoPatch})\ become $\lambda $-independent. The main
difference in comparison with Ref.\cite{Guinea} is that particle-particle
scattering leads to a $U$-dependence of the critical value $(t^{\prime }/t)_c
$ for the appearance of ferromagnetism so that $(t^{\prime }/t)_c\rightarrow
1/2$ for $U\rightarrow 0$ in qualitative agreement with the results of the $T
$-matrix approximation \cite{Fleck}. At the same time, neglecting
particle-particle scattering gives $(t^{\prime }/t)_c\simeq 0.27$
independent of $U.$ Note that the value $(t^{\prime }/t)_c$ is determined by
the condition of equal non-interacting particle-hole susceptibilities $\chi
_0({\bf 0})=\chi _0({\bf Q})$ and therefore coincides with the $``$%
mean-field'' criterion for the boundary of the ferromagnetic phase.

\subsection{Antiferromagnetic instability at small ${\bf t}^{\prime }$}

The case $t^{\prime }=0$ for fillings close to $n=1$ was intensively studied
previously within momentum-cutoff RG approaches\cite{Zanchi,Metzner}. We
plot the results of TCRG at $t^{\prime }=0$ in Fig. \ref{Nearhalf}. There is
a line of critical concentrations $n_c(U)$ such that for $n<n_c(U)$ the
extrapolated inverse susceptibility $\chi _{\text{AF}}^{-1}$ does not reach
zero for any temperature and the ground state is expected not to have
long-range antiferromagnetic order. For comparison, we also plot the result
for the critical concentrations obtained within the momentum-cutoff RG
approach, Ref. \cite{Metzner}. Both approaches give practically
indistinguishable results for the critical fillings where antiferromagnetism
disappears. Mean-field theory predicts a broader concentration range for the
stability of antiferromagnetism, see Fig. \ref{Nearhalf}. It was proven by
van Dongen \cite{Dongen} that for large space dimensionality $d\gg 1$ and $%
U\ll t$ the critical hole concentration $\delta _c=1-n_c$ is reduced in
comparison with its mean-field value $\delta _c^{\text{MF}}$ by a {\it finite%
} factor $q_d.$ To analyze wether this remains true for $d=2$ we consider
the $U$-dependence of the ratio $q(U)=\delta _c/\delta _c^{\text{MF}}.$ We
find that $q(U)$ slightly decreases with decreasing $U$ and it is saturating
at $q(0)\simeq 0.4\pm 0.025$. Surprisingly, the formal application of the
results of $1/d$ expansion in Ref. \cite{Dongen} to $d=2$ gives a close
value, $q_2=0.3.$

We have verified that within the antiferromagnetic phase at $t^{\prime }=0$
the susceptibilities at wavevectors ${\bf Q}\neq (\pi ,\pi )$ are always
smaller than the susceptibility $\chi _{\text{AF}},$ so that the tendency
towards incommensurate magnetic order is subleading in comparison with
commensurate $(\pi ,\pi )$ order. The fact that we identify commensurate AF
even away from half filling may be reconciled with the possibility that the
system develops inhomogeneous spin and charge structures (e.g. phase
separation), as was obtained in the mean-field studies of the Hubbard model
\cite{PS} and the weak-coupling results in high dimensions \cite{Dongen}.

Outside the antiferromagnetic region we find the tendency towards $d$SC as
observed previously in Refs. \cite{Metzner,SalmHon02}. The values of the
crossover temperature $T_{\text{dSC}}^{*}$ rapidly decrease away from the
antiferromagnetic phase; we show in Fig. \ref{Nearhalf} the contour lines
with $\ln (t/T_{\text{dSC}}^{*})=5,6,7.$ The contour lines with larger $T_{%
\text{dSC}}^{*}$ can not be traced within the present RG analysis: because
of strong fluctuations near the AF phase, the coupling constants reach $%
V_{\max }=18t$ before the $d$-wave susceptibility becomes large. On the
other hand, smaller $T_{\text{dSC}}^{*}$ (and correspondingly larger
deviations from half filling) are hard to treat, too, because of the
difficulties with the numerical integrations in Eqs. (\ref{dV}) since the
integrands contain sharp Fermi functions at small temperatures.

The half-filled case at different $t^{\prime }$ was investigated previously
within a mean-field analysis \cite{MFHalffil,MFHalffil1,MFHalffil2}, QMC
calculations \cite{LinHirsch,MFHalffil1}, and path-integral RG \cite{PIRG}.
Different methods predict different values of the critical interaction $U_c$
for the onset of antiferromagnetism at fixed $t^{\prime }$. In particular
for $t^{\prime }/t=0.2$ QMC results on an $8\times 8$ lattice \cite
{LinHirsch,MFHalffil1} at $T=0.25t$ yield $U_c=2.5t,$ while path-integral RG
\cite{PIRG} gives $U_c=3.4t.$ The result of the mean-field approach for the
same $t^{\prime }/t$ is $U_c=2t$ \cite{MFHalffil,MFHalffil1,MFHalffil2}. We
present our phase diagram as obtained from the many-patch RG analysis in
Fig. \ref{Halffil}; symbols show the critical values $U_c$ obtained by other
methods. As expected and in agreement with previous studies, the critical $%
U_c$ is larger than the mean-field value for all $t^{\prime }$. At the same
time, the $U_c$ result of the TCRG at $t^{\prime }/t=0.2$ is larger than
that from QMC calculations, but it agrees well with the path-integral RG
result in Ref. \cite{PIRG}.

Again, we find the tendency towards $d$SC away from the AF region; we show
in Fig. \ref{Halffil} the contour lines which correspond to $\ln (t/T_{\text{%
dSC}}^{*})=5,6,7$. Note that from the extrapolation of these data to larger $%
U$\ in the paramagnetic phase ($U<U_c$) we always find $\ln (t/T_{\text{dSC}%
}^{*})>3$ in the weak-coupling regime, i.e. a temperature regime which is
far below the accessible temperature range in QMC simulations. Therefore, it
may be difficult if not impossible to observe the corresponding
superconducting fluctuations in QMC calculations on finite lattice sizes -
at least in the weak-to intermediate coupling regime.

\subsection{Ferromagnetic instability}

Now we investigate the ferromagnetic instability, which arises for $%
t^{\prime }/t$ close to $1/2.$ We start with $t^{\prime }/t=1/2,$ when the
dispersion at the bottom of the band at small $k_x$ or $k_y$ can be expanded
as
\begin{equation}
\varepsilon _{{\bf k}}=\left\{
\begin{array}{cc}
tk_x^2(1-\cos k_y)-\mu , & k_x\ll 1 \\
tk_y^2(1-\cos k_x)-\mu , & k_y\ll 1
\end{array}
\right. ,
\end{equation}
i.e. it has extended minima along the lines $k_x=0$ and $k_y=0$ (see Fig.
\ref{Disp050}a) rather than a single minimum at the origin, as for $%
t^{\prime }/t<1/2$. This peculiar flatness of the spectrum leads to a
square-root divergence of the density of states, $\rho (\varepsilon )\propto
\varepsilon ^{-1/2}$ at the bottom of the band (Fig. \ref{Disp050}b).
Therefore in the low density limit (which is close to a vH band filling,
since $n_{VH}=0$ for $t^{\prime }/t=1/2)$, saturated ferromagnetism is
expected\cite{Fleck,Hlubina,Hlubina1}. At $t^{\prime }/t=1/2$ the $T$-matrix
approximation \cite{Hlubina} predicts rather high critical densities for the
stability of ferromagnetism, e.g. $n_c=0.57$ for $U=4t.$ For $t^{\prime
}/t=0.45$ and $U=4t$ the same approach predicts ferromagnetism for densities
$0.3<n<0.5$; the smallest value of $t^{\prime }$ at which ferromagnetism can
exist was predicted to be $(t^{\prime }/t)_c=0.43$ for $U=4t.$ The projected
QMC calculations\cite{Hlubina1} confirmed the existence of ferromagnetism
for $(t^{\prime }/t)_c\gtrsim 0.47$.

The phase diagram obtained within the TCRG approach for $t^{\prime }/t=1/2$
is shown in Fig. \ref{Flatband}. Similar to the antiferromagnetic
instability, mean-field theory overestimates the tendency to magnetic order.
The result of the $T$-matrix analysis of Ref. \cite{Hlubina} for the
critical concentration of the stability of ferromagnetism at $U=4t$ is
marked by a cross. Surprisingly, this result is very close to the result of
the RG approach. Similar to Ref. \cite{Dongen} one may introduce the
quantity $q_{\text{F}}(U)=n_c/n_c^{\text{MF}}$ to measure the deviation from
the mean-field result at $t^{\prime }/t=1/2.$ The analysis of the data shows
that $q_{\text{F}}(U)$ slightly increases with decreasing $U$ and $q_{\text{F%
}}(U\rightarrow 0)\simeq 0.8$.

We have also explored the possibility for triplet ($p$-wave) pairing in the
vicinity of the ferromagnetic phase. Although the $p$-wave pairing
susceptibility is dominant in this region, a conclusive low-temperature
extrapolation for the inverse susceptibility is not possible. Therefore it
is not clear whether a finite crossover temperature $T_{\text{pSC}}^{*}$
exists. In any case the possible values for $T_{\text{pSC}}^{*}$ must be
significantly smaller than the crossover temperatures for $d$-wave
superconductivity. The region where $\ln (t/T_X)<8$ is shown in Fig. \ref
{Flatband}, too. The growing of the vertices near the ferromagnetic phase
results from the triplet $p$-wave superconducting fluctuations, but,
unfortunately, the smallness of the temperature crossover scale, which is
far below the range of applicability of the TCRG method prevents a safe
conclusion about the possibility of a $p$-wave superconducting ground state.

Now we consider the case $t^{\prime }/t<1/2,$ which is very different from
the above-discussed case $t^{\prime }/t=1/2.$ The square-root divergence of
the density of states is replaced by a logarithmical divergence at the
energy of the vHS, $\rho (\varepsilon )\propto \ln (t/\varepsilon )$, while
the density of states is finite at the lower band edge (see Fig. \ref
{Disp045}). The phase diagram for $t^{\prime }/t=0.45$ is presented in Fig.
\ref{Nearflat}, where we again mark by cross the result of the $T$-matrix
approximation. The ferromagnetic region substantially shrinks with
decreasing $t^{\prime }$: it reduces to a narrow density window around the
vH band filling $n_{VH}=0.465$ (the corresponding critical densities are
almost symmetrical around $n_{VH}$ so that only the region $n>n_{VH}$ is
shown). Nevertheless, the ferromagnetic region is wider than in the $T$%
-matrix approximation. The same tendency is reflected in the RG result for
the critical value $(t^{\prime }/t)_c\approx 0.3$ for the disappearance of
ferromagnetism at $U=4t$, which is much lower than the result of the $T$%
-matrix approximation cited above, $(t^{\prime }/t)_c=0.43$. As well as for
the case $t^{\prime }=1/2$ we also find an increasing triplet
superconducting susceptibility away from the ferromagnetic phase, while the
possible corresponding crossover temperatures $T_{\text{pSC}}^{*}$ remain
undetectably small.

\section{Summary and conclusions}

We have considered the phase diagrams of the $t$-$t^{\prime }$ Hubbard model
within two- and many-patch RG approaches as shown in Figs. \ref{vanHove},
\ref{Nearhalf}, \ref{Halffil}, \ref{Flatband}, and \ref{Nearflat}.
Instabilities towards antiferro- or ferromagnetic order as well as to
singlet $d$-wave superconductivity are identified in different parameter
regimes. Near the ferromagnetic region the $p$-wave superconducting
susceptibility is enhanced, but a conclusion about a possible triplet
superconducting ground state remains elusive.

At small $t^{\prime }$ and vH band fillings the antiferromagnetic
instability dominates. With increasing $t^{\prime }$ antiferromagnetism is
replaced by $d$-wave superconductivity. At larger $t^{\prime }/t$
ferromagnetism becomes the leading instability. The tendency towards $d$%
--wave superconductivity decreases with increasing $U$ while
antiferromagnetism is enhanced. We found that the two-patch approach
predicts correctly the boundary of the ferromagnetic phase at vH band
fillings, while it fails to reproduce correctly the boundary between
antiferromagnetic and superconducting phases at small $t^{\prime }$ where
nearly nesting effects become important.

Antiferromagnetism at small $t^{\prime }$ and ferromagnetism at $t^{\prime
}/t=1/2$ exist in broad density ranges around vH band fillings; the
antiferromagnetism remains commensurate in the part of the phase diagram
where the long-range ordered ground state is expected. The density ranges
for magnetic order, found from TCRG are substantially narrower ($2.5$ times
for the AF instability and $1.3$ times for the F instability at small $U$)
than the corresponding mean-field results.

At half-filling at different $t^{\prime }$ we find the critical interaction
strengths for the antiferromagnetic instability. From the present analysis
we can not argue, wether the antiferromagnetic state we find is metallic or
insulating. It was proposed \cite{MFHalffil1} that at nonzero $t^{\prime }$
there is a finite interaction range $U_c<U$ $<U_c^{\prime }$ for metallic
antiferromagnetism, at $U>U_c^{\prime }$ it is replaced by the insulating AF
state. On the other hand, the existence of a paramagnetic insulating state
at larger $t^{\prime }$ was conjectured in Ref. \cite{PIRG}. Discriminating
between these possibilities requires the calculation of the conductivity and
the Drude weight, for which it is necessary to retain the frequency
dependence of the vertices.

The boundary of ferromagnetism at $t^{\prime }=t/2$ found from TCRG is
surprisingly close to the $T$-matrix approximation result in Ref. \cite
{Hlubina} at $U=4t$ , although the corresponding filling is not small and
possibly outside the region of the validity of the $T$-matrix approximation.
In the vicinity of antiferro- and ferromagnetic phases we found regions with
enhanced $d$-wave and $p$-wave superconductivity, respectively. Not too
close to the antiferromagnetic phase the crossover temperatures for $d$-wave
superconductivity into the corresponding renormalized classical regime with
exponentially large correlation length can be estimated from the
extrapolation to low temperatures of RG data for the order-parameter
susceptibilities. At the same time, triplet $p$-wave superconductivity in
the vicinity of the ferromagnetic phase possibly has much smaller crossover
temperatures $T_{\text{pSC}}^{*}$ which can not be determined safely from
the present RG analysis.

The ferromagnetic phase substantially shrinks for $t^{\prime }/t<1/2$ and
the difference to the mean-field result increases. For this case, the $T$%
-matrix approach underestimates the tendency towards ferromagnetism. The
tendency towards triplet $p$-wave superconductivity in the vicinity of the
ferromagnetic phase persists, although its associated temperature crossover
scale remains very low - significantly lower than for $d$-wave
superconductivity.

It remains an open issue, how the above results change, when the two-loop
corrections to the RG equations are taken into account, and how the
electronic self-energy evolves in the vicinity of magnetic or
superconducting instabilities. Another interesting issue for future work
remains the question whether inhomogenous spin and charge structures are
possible near half-filling and whether the Pomeranchuk instability may
become the leading instability for anisotropic extensions of the 2D Hubbard
model; if it does it is natural to connect this tendency to the stripe
pattern formation in rare-earth doped La$_{2-x}$Sr$_x$CuO$_4.$ Furthermore,
the tendency towards $p$-wave superconductivity near the ferromagnetic phase
of the $t$-$t^{\prime }$ model suggests a possible route fo future
investigations of the origin of triplet pairing in Sr$_2$RuO$_4.$

\section*{Acknowledgements}

We are grateful to W. Metzner, G. Uhrig, and M. I. Katsnelson for insightful
discussions. This work was supported by the Deutsche Forschungsgemeinschaft
through SFB 484.

\section*{Appendix. The temperature crossover to the renormalized classical
regime}

In this Appendix we discuss how the temperature dependence of the magnetic
susceptibilities changes at the crossover to the renormalized-classical
regime. As a first example, we consider the 2D ferro- and antiferromagnetic
Heisenberg models
\begin{equation}
H=\pm J\sum_{\langle ij\rangle }{\bf S}_i\cdot {\bf S}_j
\end{equation}
(plus corresponds to the antiferro-, minus to the ferromagnet, $J>0$). The
susceptibility (staggered susceptibility in the AF case) at high
temperatures $T\gg J$ obeys the Curie law
\begin{equation}
\chi _{\text{F,AF}}=\frac CT
\end{equation}
where $C=JS(S+1)/3$. On the other hand, at temperatures $T\ll J$ it was
found from the RG analysis of the 2D nonlinear sigma model\cite{CHN,Kopietz}
\begin{equation}
\chi _{\text{F,AF}}=C^{\prime }\frac T{J^2}\xi ^2
\end{equation}
where
\begin{equation}
\xi =C_\xi \left\{
\begin{array}{cc}
\exp (2\pi \rho _s/T) & \text{AF} \\
(T/J)^{1/2}\exp (2\pi JS/T) & \text{F}
\end{array}
\right.
\end{equation}
is the correlation length, $C_\xi ,C^{\prime }$ are temperature-independent
prefactors, and $\rho _s$ is the zero-temperature value of the spin
stiffness, which is proportional to the ground-state (sublattice)
magnetization $\overline{S}_0.$ Therefore, below the crossover temperature $%
T^{*}\sim 2\pi J\overline{S}_0$ ($\overline{S}_0=S$ for a ferromagnet) the
susceptibility becomes exponentially large.

Similar results can be obtained for the Hubbard model within the
two-particle self-consistent approach \cite{Vilk}. We have
\begin{equation}
\chi _{\text{F,AF}}=\frac{\chi _{{\bf Q}}^0(T)}{1-U_{sp}(T)\chi _{{\bf Q}%
}^0(T)}
\end{equation}
where ${\bf Q=}(\pi ,\pi )$ in the AF case, ${\bf Q=0}$ in the F case, and $%
\chi _{{\bf Q}}^0(T)=\chi _0({\bf Q},0,T)$ is the zero-frequency limit of
the bare dynamic susceptibility
\begin{equation}
\chi _0({\bf q},i\omega _n,T)=\sum_{{\bf k}}\frac{f_{{\bf k}}-f_{{\bf k+q}}}{%
i\omega _n-\varepsilon _{{\bf k}}+\varepsilon _{{\bf k+q}}}.
\end{equation}
The effective interaction $U_{sp}(T)$ satisfies the self-consistent equation
\begin{equation}
2n-n^2\frac{U_{sp}(T)}U=4T\sum_{{\bf q,}i\omega _n}\frac{\chi _0({\bf q}%
,i\omega _n,T)}{1-U_{sp}(T)\chi _0({\bf q},i\omega _n,T)}.  \label{TPSC}
\end{equation}
At high temperatures $T\gg t$ we have $U_{sp}(T)\simeq U$ and
\begin{equation}
\chi _{\text{F,AF}}\simeq \chi _{{\bf Q}}^0(T)\simeq \frac 1{4T}.
\end{equation}
At small temperatures the denominator in Eq. (\ref{TPSC}) can be expanded
for wavevectors ${\bf q}$ close to ${\bf Q}$ (cf. Ref. \cite{Vilk}) and one
obtains
\begin{equation}
\overline{\sigma }_0^2=\frac{2T\chi _0}{AN}\sum_{{\bf q}}\frac 1{\xi ^{-2}+(%
{\bf q-Q)}^2}  \label{Ess}
\end{equation}
where $\chi _0=\chi _0({\bf Q},0,0),$ $A=\nabla ^2\chi _0({\bf q},0,0)|_{%
{\bf q}={\bf Q}},$ and $\overline{\sigma }_0^2=n-n^2U_{sp}(0)/2-{\cal P}/2$ (%
${\cal P}$ is the zero-temperature value of the r.h.s. of Eq. (\ref{TPSC})),
and $\xi ^{-2}=[1-U_{sp}(T)\chi _0({\bf Q,}0,T)]/A$. Momentum integration in
Eq. (\ref{Ess}) leads to
\begin{equation}
\xi =C_\xi \exp \{nA\overline{\sigma }_0^2/2T\chi _0\}
\end{equation}
and
\[
\chi _{\text{F,AF}}=\chi _0\xi ^2.
\]
In this case the crossover temperatures $T_{\text{F},\text{AF}}^{*}\sim t%
\overline{\sigma }_0^2.$ Therefore these examples show that
\[
\chi _{\text{F,AF}}=\left\{
\begin{array}{cc}
C/T & T\gg T^{*} \\
C^{\prime }\xi ^2 & T\ll T^{*}
\end{array}
\right.
\]
and the correlation length $\xi \propto \exp ({\cal A}T^{*}/T)$ is
exponentially large in the low-temperature, renormalized classical regime$.$

\newpage
\

${\sc Figure\ captions}$

\begin{enumerate}
\item  The dependence of the van Hove band filling on $t^{\prime }/t$.

\item  The Fermi surface at van Hove band fillings: $t^{\prime }=0$ and $n=1$
(solid line), $t^{\prime }/t=0.1$ and $n=0.92$ (long-dashed line),$\ $and $%
t^{\prime }/t=0.3$ and $n=0.72$ (short-dashed line), $A$ and $B$ are van
Hove points.

\item  The division of momentum space into patches in the two-patch
approach. Regions I contain the momenta closest to the vH singularity points
${\bf k}_A$ and ${\bf k}_B,$ regions II contain the momenta which are close
to the FS but far from vH singularities and regions III contain the momenta
far from both, FS and vH singularities.

\item  The four types of vertices considered in the two-patch approach: (a)
and (b) correspond to exchange and direct scattering between different vH
singularities respectively, (c) umklapp scattering, (d) intrapatch
scattering. The incoming and outgoing momenta with equal spin are connected
by solid lines inside the vertices.

\item  a) Scaling behavior of the coupling constants at $t^{\prime }/t=0.1$.
The solid line corresponds to $g_1,$ dashed line to $g_2,$ dash-dotted line
to $g_3,$ dotted line to $g_4.$ b) Scaling behavior of the susceptibilities
at $t^{\prime }/t=0.1.$ Solid line corresponds to the AF, dashed line to the
dSC, and dotted line to the F susceptibility. The interaction strength is $%
U=2t.$ \label{g1}

\item  The same as in Fig. \ref{g1} for $t^{\prime }/t=0.45$. \label{g2}

\item  Diagrammatic representation for the many-patch RG equations, Eq. (\ref
{dV}). Lines drawn through the vertices show the direction of spin
conservation. Diagrams are drawn in the same order as the respective terms
in Eq. (\ref{dV}). The cutting dash at the propagator lines means the
derivative with respect to $T$ (for brevity we indicate only the derivative
of one of the propagators, the same diagrams with derivatives of another
propagator are included as well).\label{Diagram}

\item  Phase diagram at vH band fillings as obtained from two- and
many-patch RG analyses. Solid lines correspond to the phase boundaries
obtained within the two-patch RG analysis. The symbols show the results of
the many-patch RG approach: closed circles correspond to AF, open circles to
incommensurate ($\pi ,\pi -\delta $) order, diamonds to dSC, and triangles
to the F phase. Long-dashed line is the boundary of ferromagnetic phase
obtained in Ref. \cite{Guinea} (see text). \label{vanHove}

\item  Phase diagram for $t^{\prime }=0$. The dashed line is the mean-field
phase boundary between antiferromagnetic and paramagnetic phases, the solid
line is the boundary of the antiferromagnetic phase obtained from the
temperature-cutoff many-patch RG approach. The corresponding result of Ref.
\cite{Metzner} for the boundary of the antiferromagnetic phase is shown by
dotted line. The dot-dashed, dot-dot-dashed, and dot-dot-dot-dashed lines
are contour lines for the $d$-wave superconductiving crossover temperature
into renormalized classical regime $T_{\text{dSC}}^{*}=e^{-5}t,e^{-6}t,$ and
$e^{-7}t$ respectively (see text). The inset shows the phase diagram in $\mu
$-$U$ coordinates.\label{Nearhalf}

\item  Phase diagram at $n=1$ (half-filled case). The notations are the same
as in Fig. \ref{Nearhalf}. The cross corresponds to the critical $U_c$ for
the stability of the antiferromagnetic phase at $t^{\prime }/t=0.2$ as
obtained from QMC calculations \cite{LinHirsch}, the star marks the result
of the path-integral RG approach for $U_c$ \cite{PIRG}.\label{Halffil}

\item  The electronic dispersion (a) and non-interacting density of states
(b) at $t^{\prime }/t=1/2.$\label{Disp050}

\item  The phase diagram for $t^{\prime }/t=1/2.$ The long-dashed line MF(F)
is the mean-field phase boundary between ferromagnetic and paramagnetic
phases, the dot-dashed line MF(S) is the mean-field result for the boundary
of saturated ferromagnetism. The solid line is the boundary of ferromagnetic
phase obtained from the temperature-cutoff many-patch RG approach, the
short-dashed line is the contour line above which the maximal vertex reach $%
V_{\max }=18t$ at the temperature $T_X>e^{-8}t.$ The $T$-matrix phase
boundary for the ferromagnetic phase \cite{Hlubina} at $U=4t$ is marked by
cross. The inset shows the phase diagram in $\mu $-$U$ coordinates.\label
{Flatband}, pSC marks the region where the triplet superconducting
susceptibility is dominating.

\item  The non-interacting density of states for $t^{\prime }=0.45t.$\label
{Disp045}

\item  The phase diagram for $t^{\prime }/t=0.45.$ The notations are the
same as in Fig. \ref{Flatband}.\label{Nearflat}
\end{enumerate}

\newpage

\psfig{file=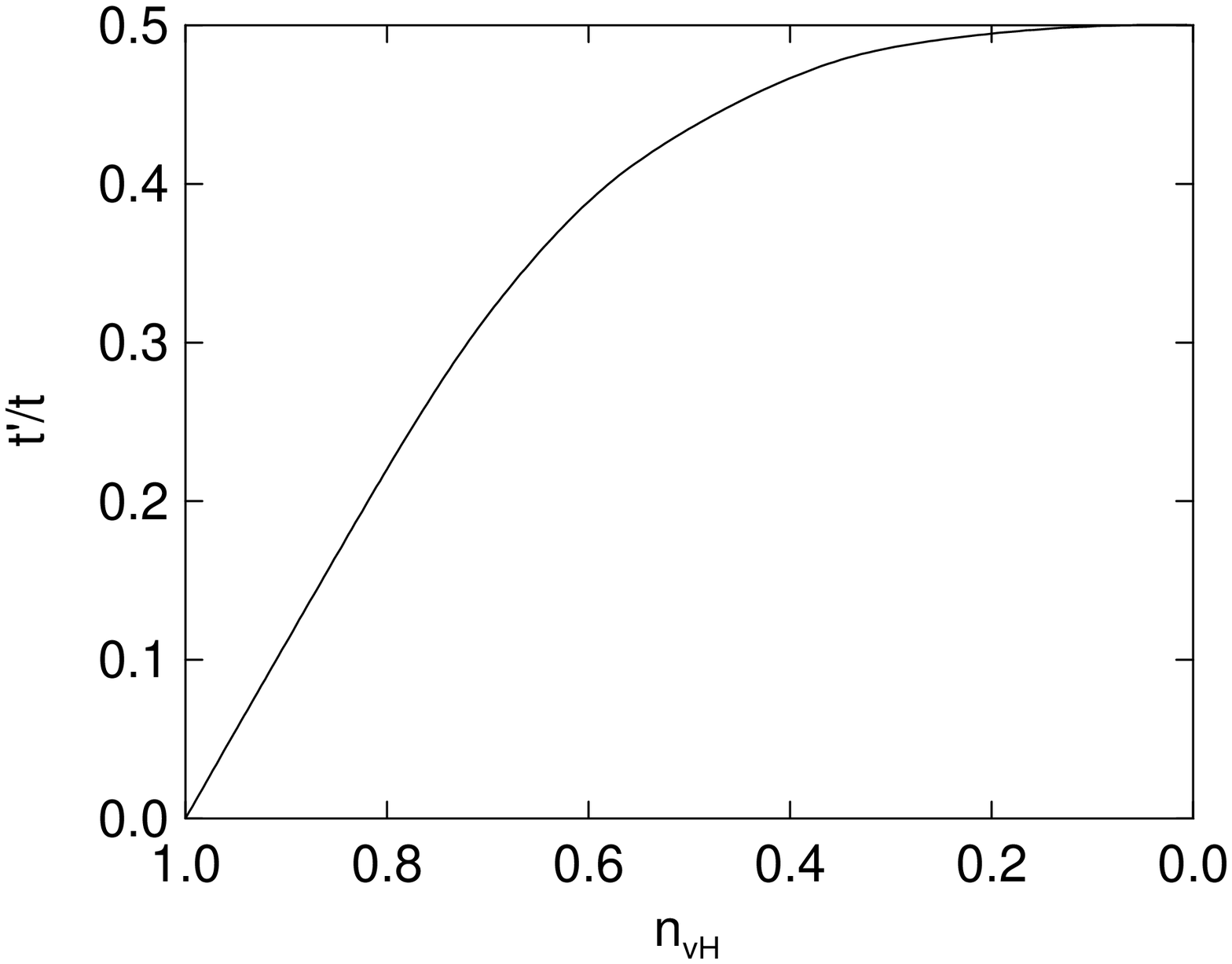}

\newpage

\psfig{file=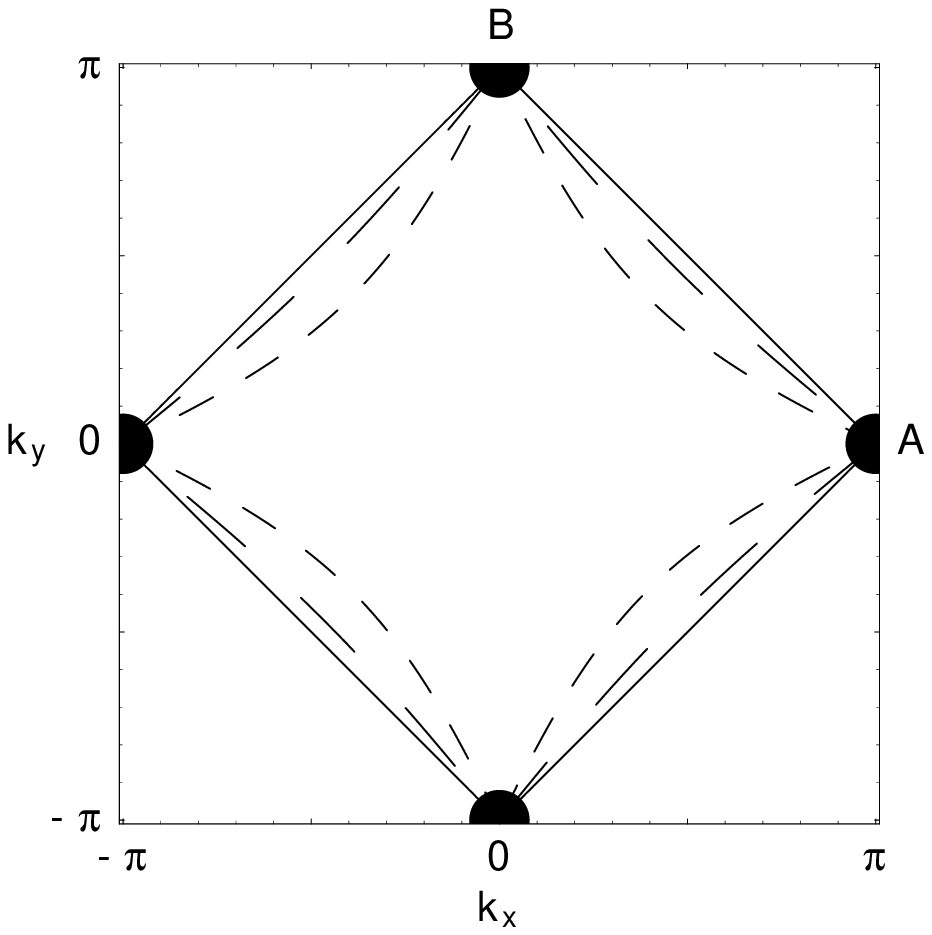}

\newpage

\psfig{file=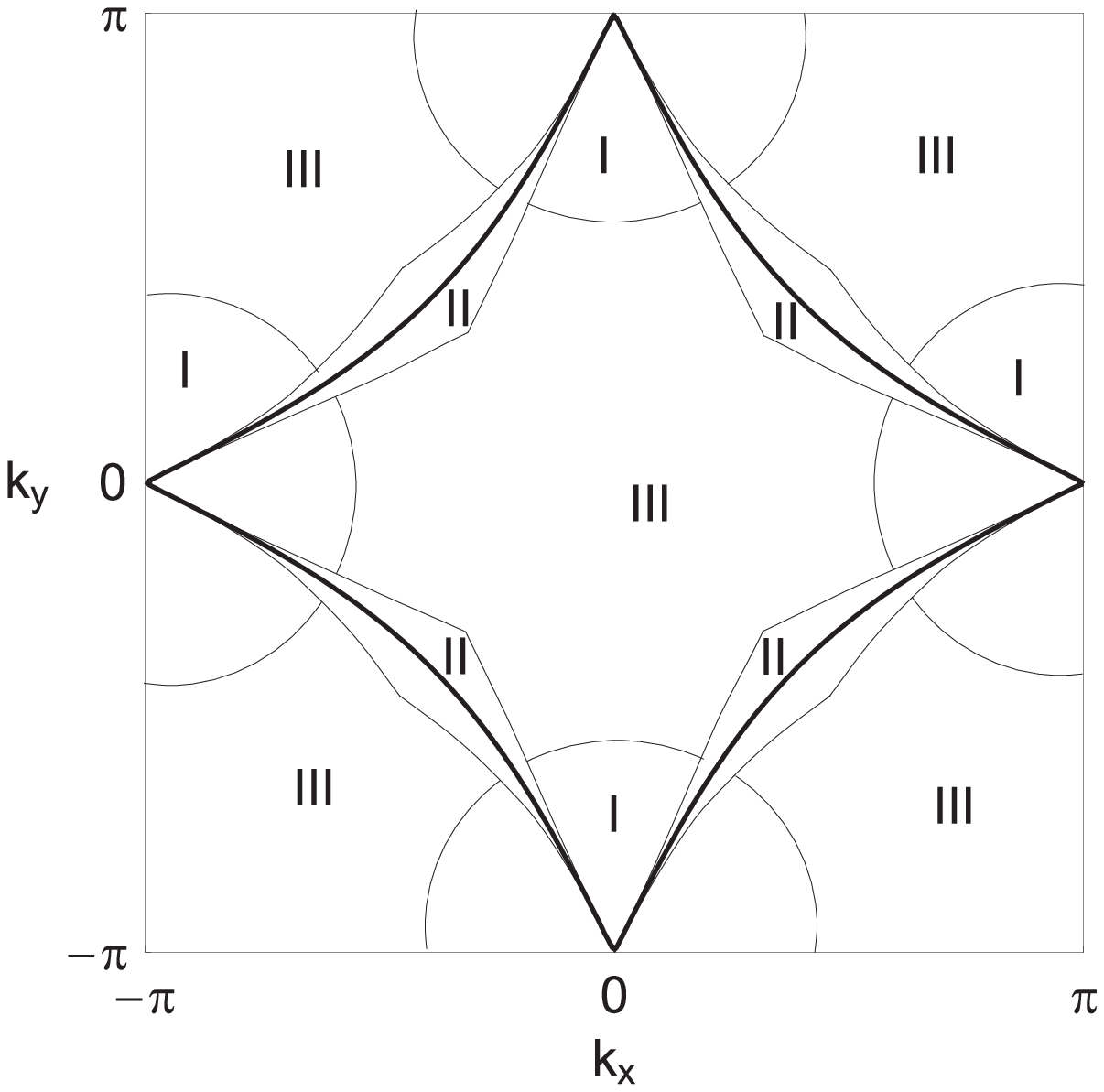}
\vspace{2cm}
\psfig{file=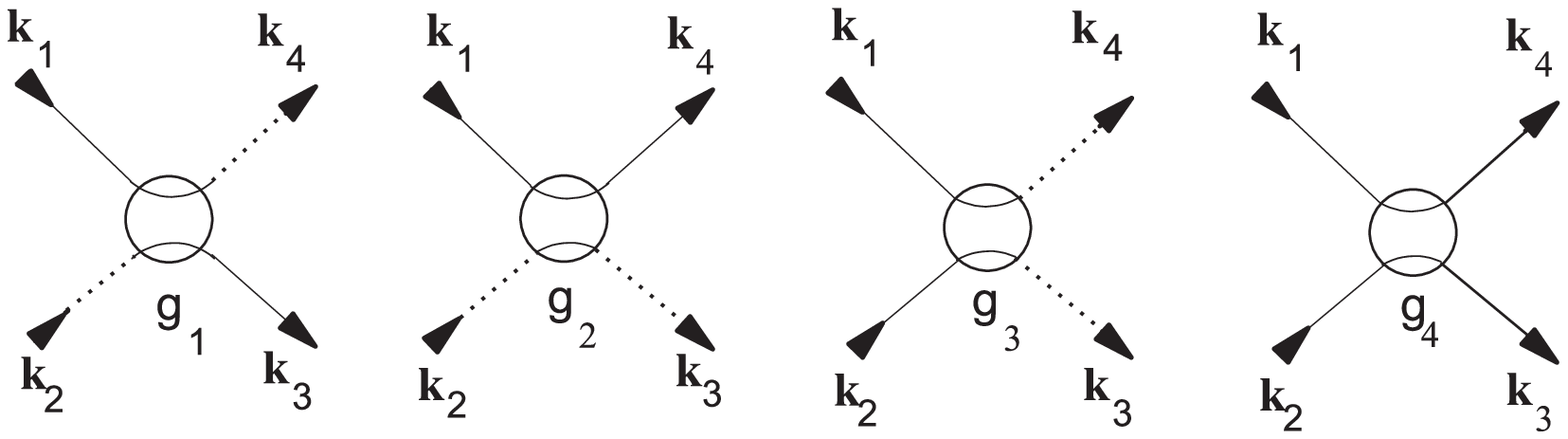}

\newpage

\psfig{file=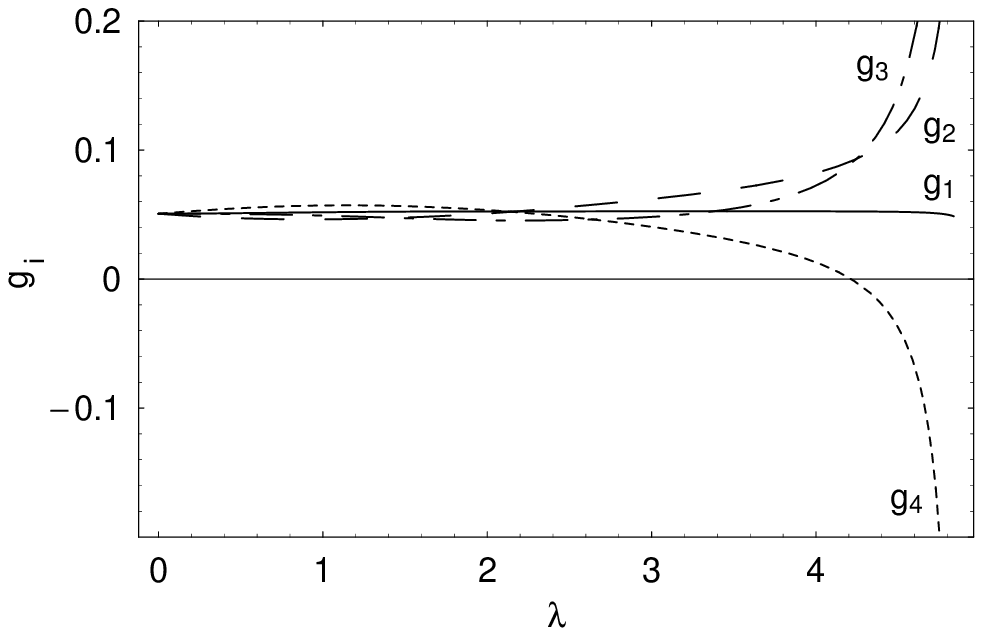}
\vspace{0.5cm}
\psfig{file=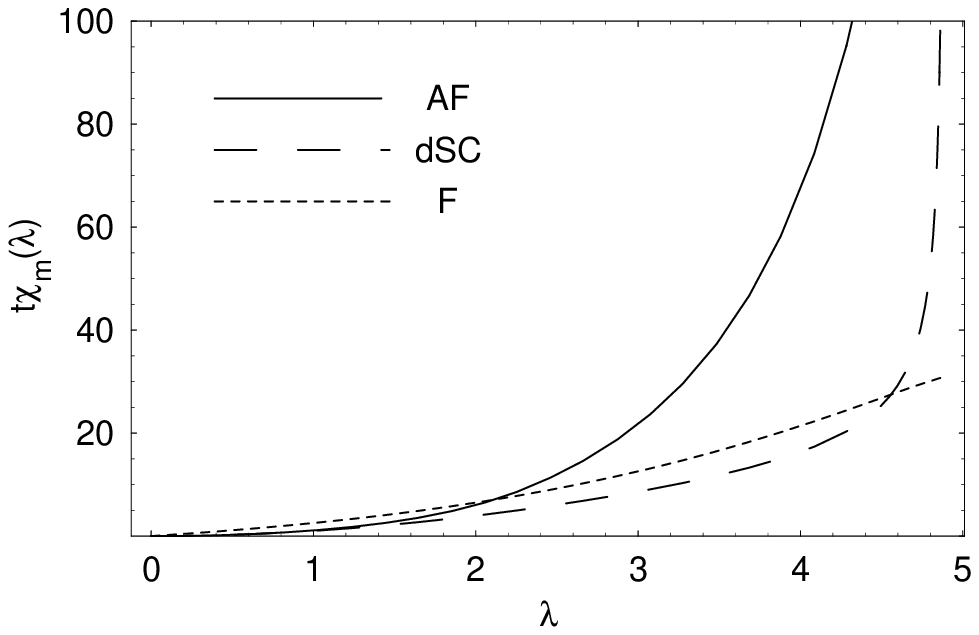}

\newpage

\psfig{file=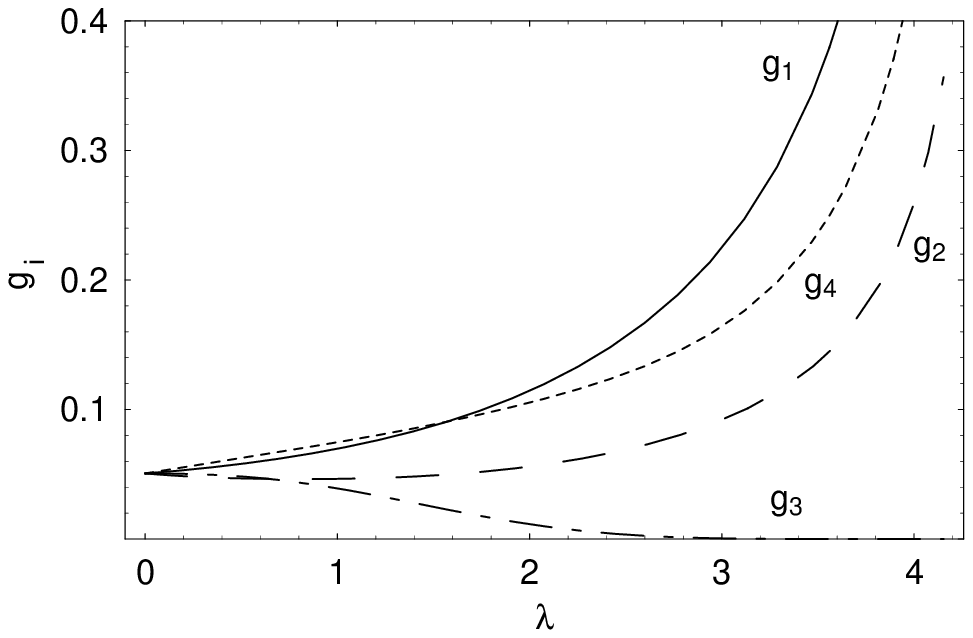}
\vspace{0.5cm}
\psfig{file=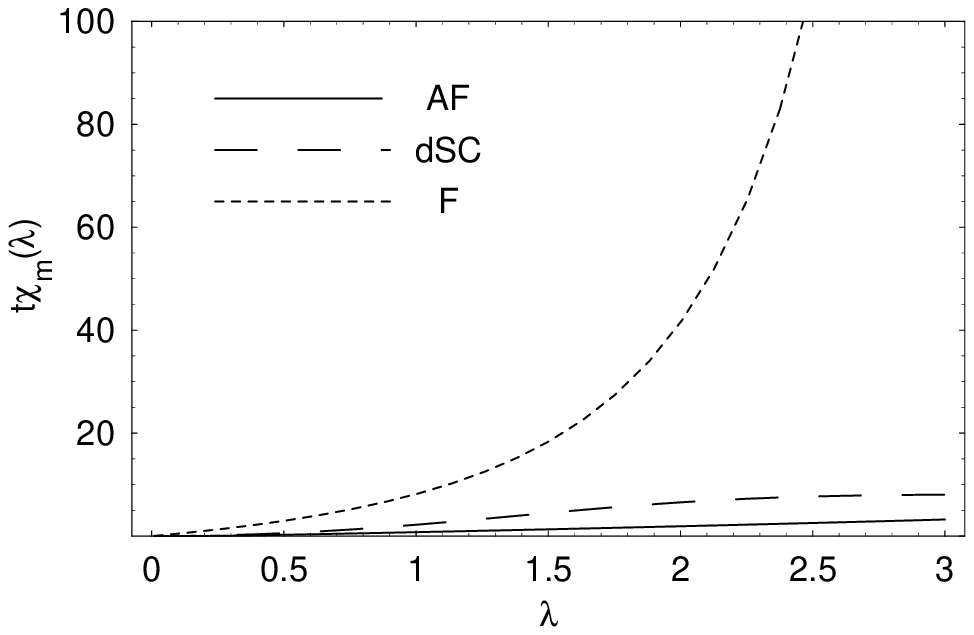}

\newpage

\psfig{file=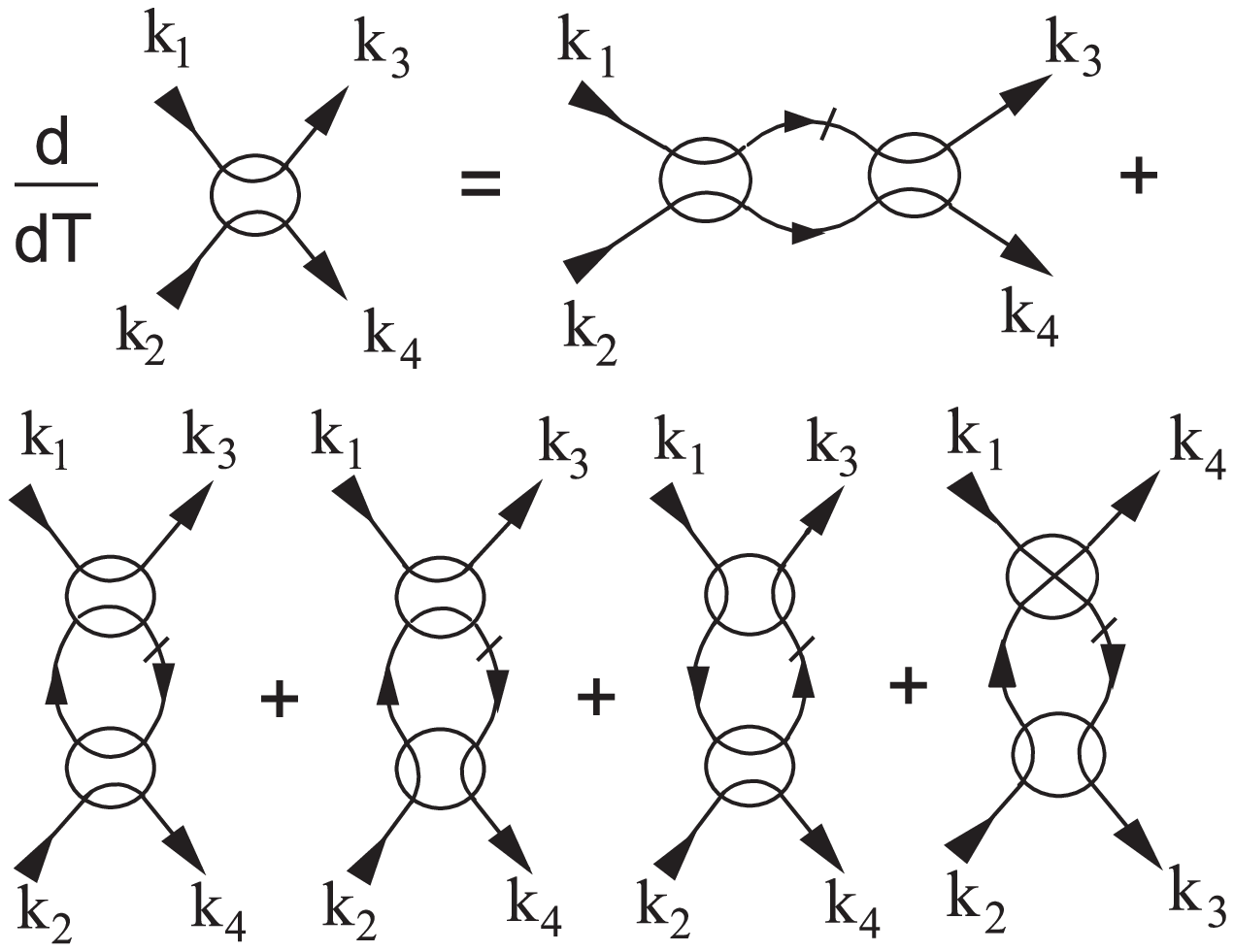}

\newpage

\psfig{file=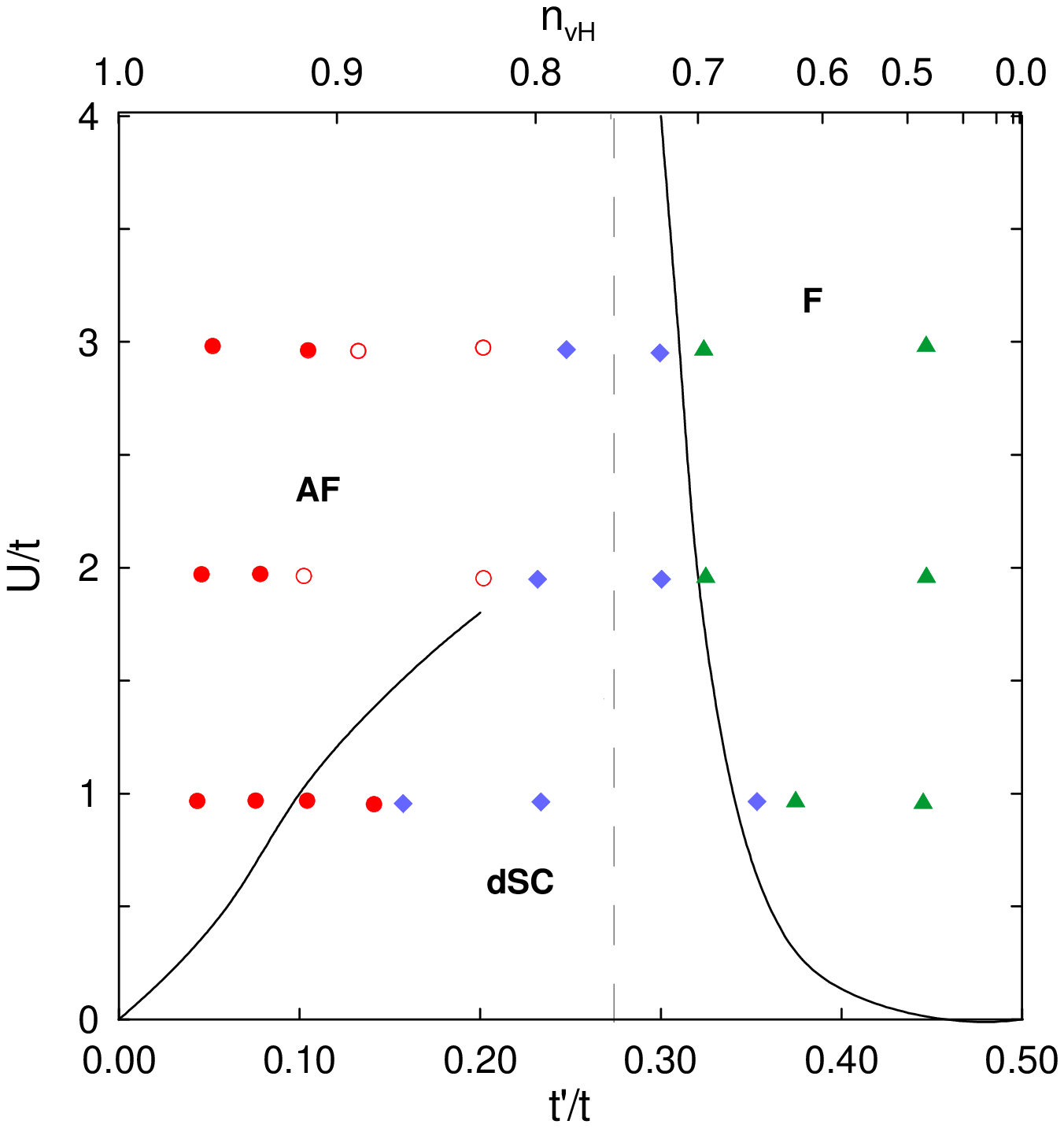}

\newpage

\psfig{file=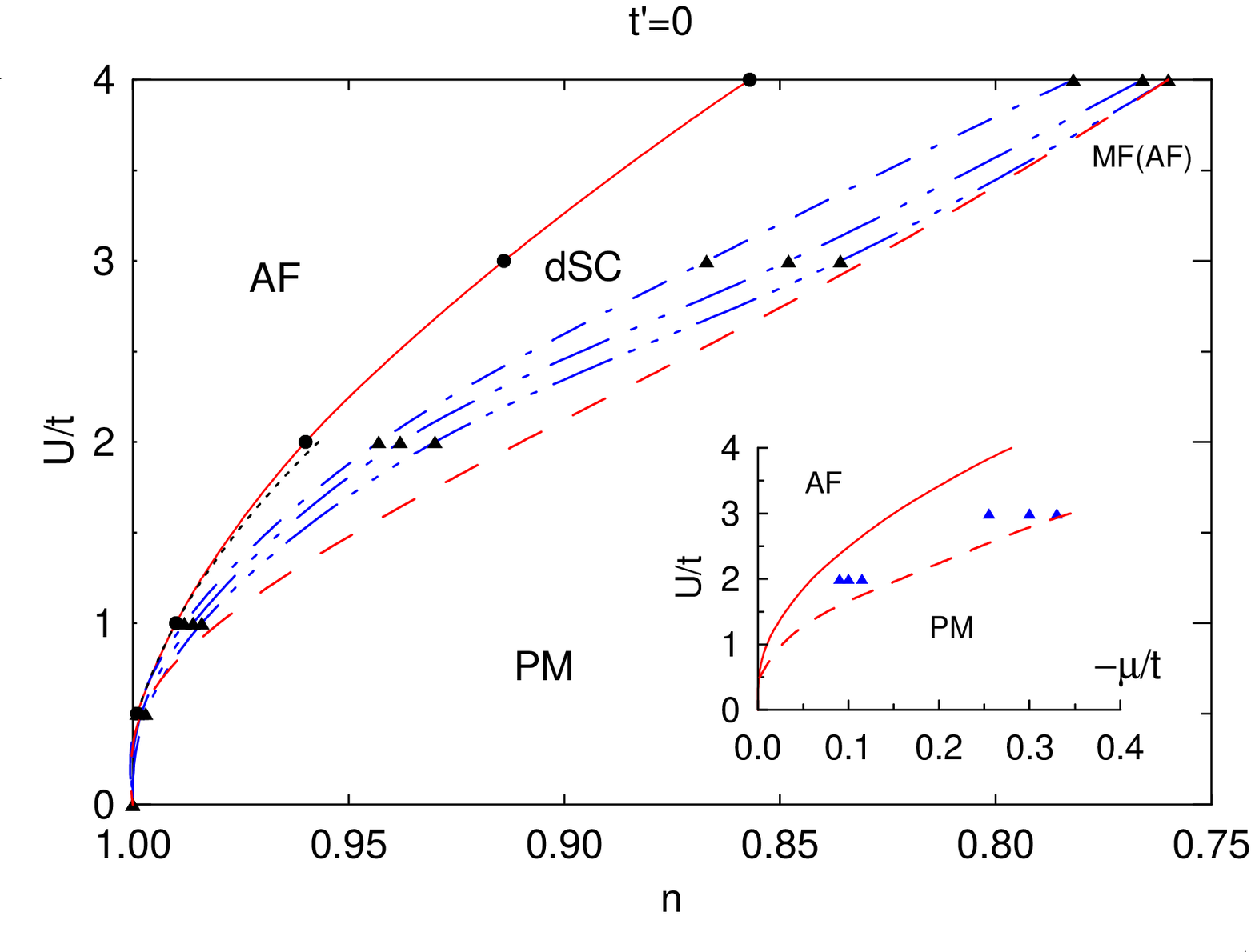}

\newpage

\psfig{file=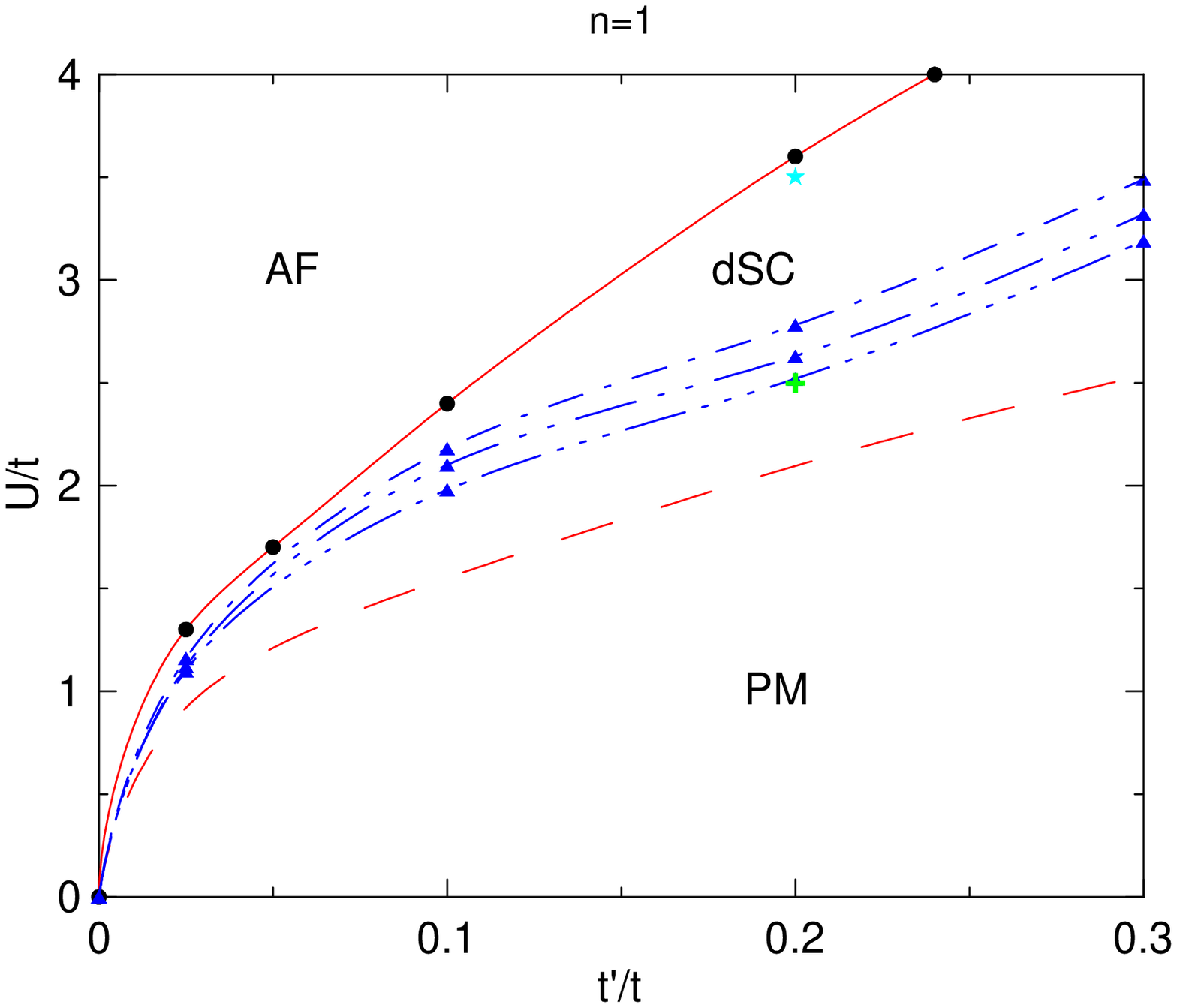}

\newpage

\psfig{file=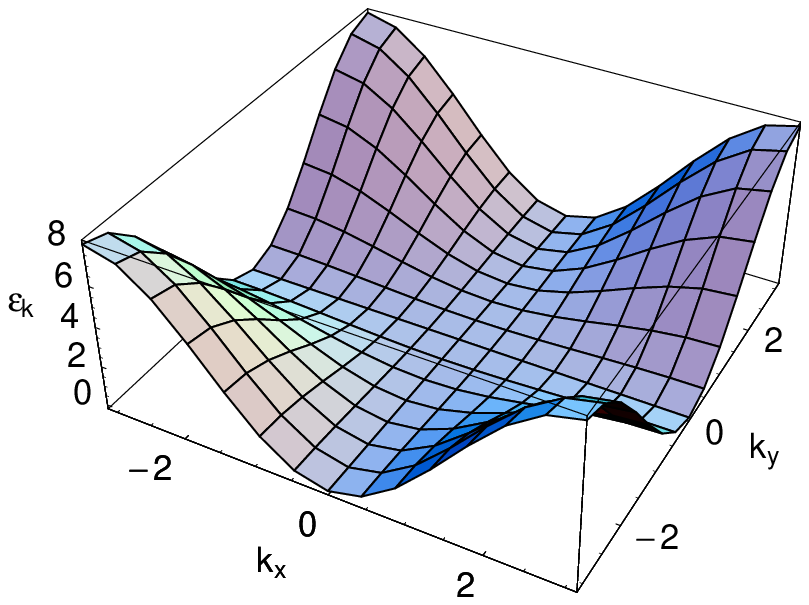}
\vspace{0.5cm}
\psfig{file=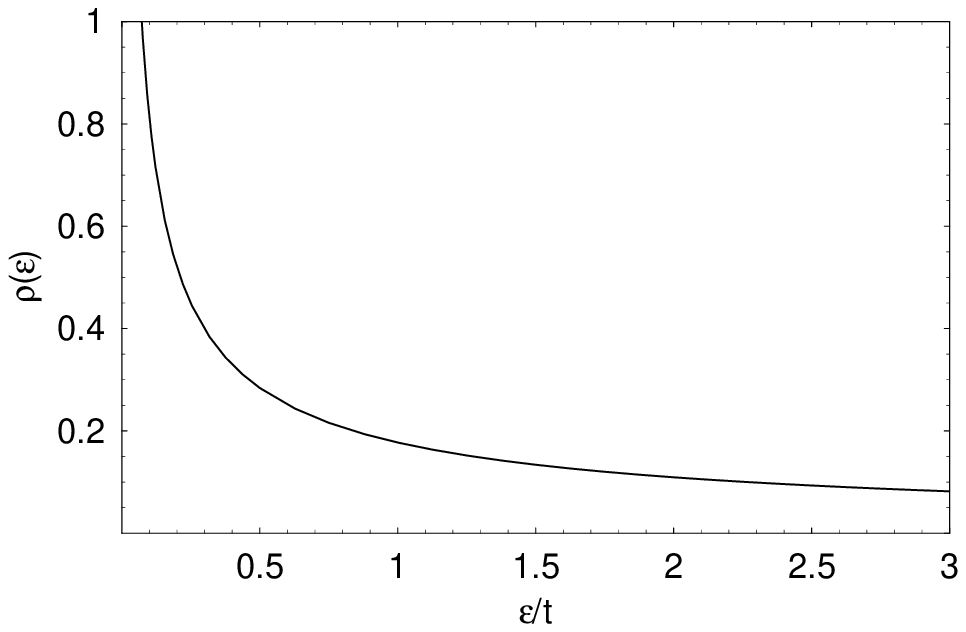}

\newpage

\psfig{file=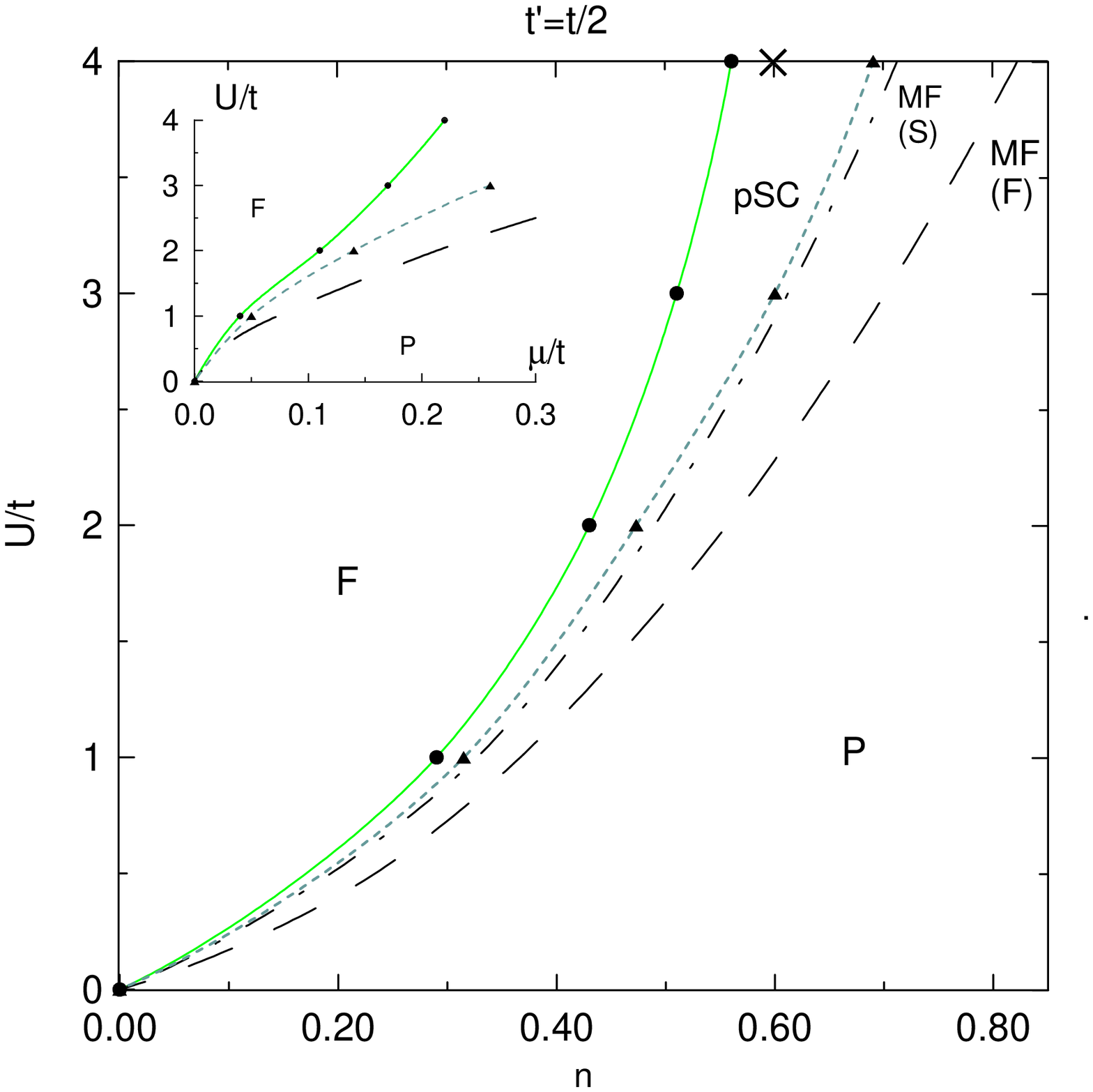}

\newpage

\psfig{file=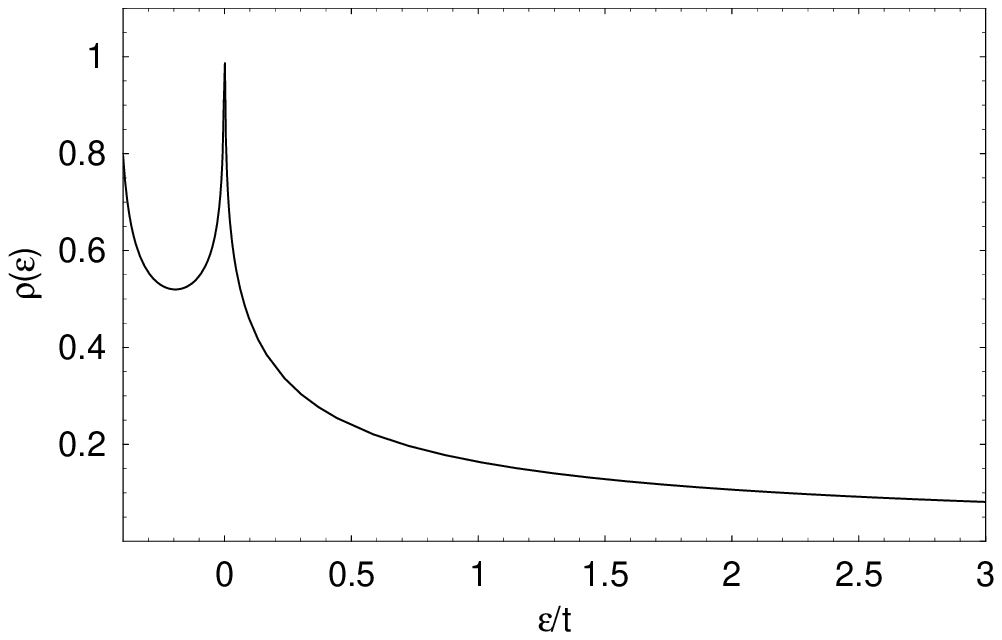}

\newpage

\psfig{file=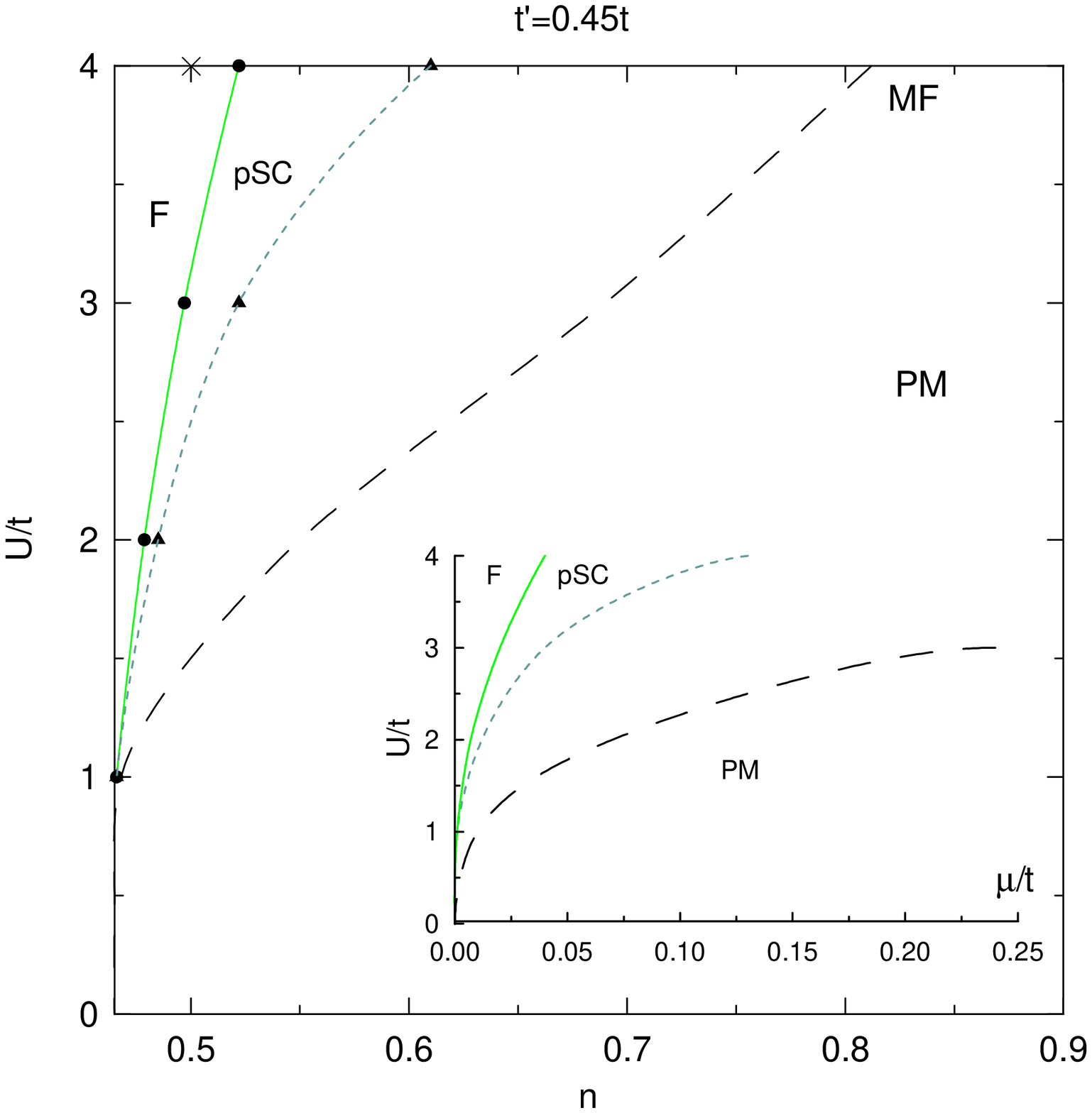}


\begin{references}
\bibitem{Scalapino}  D. J. Scalapino, Phys. Rep. {\bf 251,} 1 (1994); J. Low
Temp. Phys. {\bf 117,} 179 (1999).

\bibitem{BSW}  N. E. Bickers, D. J. Scalapino, and S. R. White, Phys. Rev.
Lett. {\bf 62,} 961 (1989).

\bibitem{Zhang}  S. C. Zhang, Science {\bf 275,} 1089 (1997); E. Demler and
S. C. Zhang, Nature (London) {\bf 396,} 733 (1998).

\bibitem{Pines}  J. Schmalian, D. Pines, and B. Stojkovic, Phys. Rev. Lett.
{\bf 80,} 3839 (1998).

\bibitem{Chubukov}  A. Chubukov, D. Pines, and B. Stojkovic, J. Phys.:
Condens. Matter {\bf 8}, 10017 (1996); A. Chubukov and D. Morr, Phys. Rep.
{\bf 288}, 355 (1997); A. Abanov and A. Chubukov, Phys. Rev. Lett. {\bf 84},
5608 (2000).

\bibitem{PTReview}  Y. Maeno, T. M. Rice, and M. Sigrist, Physics Today {\bf %
54}, 42 (2001).

\bibitem{Mazin}  I. I. Mazin and D. J. Singh, Phys. Rev. Lett. {\bf 79}, 733
(1997); ibid. {\bf 82}, 4324 (1999).

\bibitem{Murakami}  S. Murakami, N. Nagaosa, and M. Sigrist, Phys. Rev.
Lett. {\bf 82}, 2939 (1999).

\bibitem{Maeno}  Y. Sidis, M. Braden, P. Bourges, B. Hennion, S. Nishizaki,
Y. Maeno, and Y. Mori, Phys. Rev. Lett. {\bf 83}, 3320 (1999).

\bibitem{ED}  N. Kikugawa and Y. Maeno, cond-mat/0211248 (unpublished).

\bibitem{Maeno1}  F. Nakamura, T. Goko, M. Ito, T. Fujita, S. Nakatsuji, H.
Fukazawa, Y. Maeno, P. Alireza, D. Forsythe, and S. R. Julian, Phys. Rev. B
{\bf 65}, 220402(R) (2002).

\bibitem{ARPES}  A. Ino, C. Kim, M. Nakamura, T. Yoshida, T. Mizokawa, A.
Fujimori, Z.-X. Shen, T. Kakeshita, H. Eisaki, and S. Uchida, Phys. Rev. B
{\bf 65}, 094504 (2002).

\bibitem{ARPES1}  P.V. Bogdanov, A. Lanzara, X.J. Zhou, S.A. Kellar, D.L.
Feng, E.D. Lu, H. Eisaki, J.-I. Shimoyama, K. Kishio, Z. Hussain, and Z. X.
Shen, Phys. Rev. B {\bf 64}, 180505 (2001).

\bibitem{ARPES2}  D.L. Feng, C. Kim, H. Eisaki, D.H. Lu, A. Damascelli, K.M.
Shen, F. Ronning, N.P. Armitage, N. Kaneko, M. Greven, J. Shimoyama, K.
Kishio, R. Yoshizaki, G.D. Gu, and Z.-X. Shen, Phys. Rev. B {\bf 65},
220501(R) (2002).

\bibitem{Damascelli}  A. Damascelli, D. H. Lu, K. M. Shen, N. P. Armitage,
F. Ronning, D. L. Feng, C. Kim, Z.-X. Shen, T. Kimura, Y. Tokura, T.
Tsukuba, Q. Mao, and Y. Maeno, Phys. Rev. Lett. {\bf 85}, 5194 (2000).

\bibitem{t'/t}  T. Tohyama and S. Maekawa, Supercond. Sci. Tech. {\bf 13},
R17 (2000).

\bibitem{SrBand}  T. Oguchi, Phys. Rev. B {\bf 51}, 1385 (1995); D. J.
Singh, Phys. Rev. B {\bf 52}, 1358 (1995).

\bibitem{ARS}  D. F. Agterberg, T. M. Rice, and M. Sigrist, Phys. Rev. Lett.
{\bf 78}, 3374 (1997).

\bibitem{LinHirsch}  H. Q. Lin and J. E. Hirsch, Phys. Rev. B {\bf 35}, 3359
(1987)

\bibitem{Santos}  R. R. Santos, Phys. Rev. B {\bf 39}, 7259 (1989).

\bibitem{Kampf}  B. Normand and A. P. Kampf, Phys. Rev. B {\bf 65}, 020509
(2002).

\bibitem{Tasaki}  H. Tasaki, Phys. Rev. Lett. {\bf 69}, 1608 (1992); A.
Mielke and H. Tasaki, Commun. Math. Phys. {\bf 158}, 341 (1993).

\bibitem{Fleck}  M. Fleck, A. Oles, and L. Hedin, Phys. Rev. B {\bf 56},
3159 (1997).

\bibitem{Hlubina}  R. Hlubina, Phys. Rev. B {\bf 59}, 9600 (1999)

\bibitem{Hlubina1}  R. Hlubina, S. Sorella, and F. Guinea, Phys. Rev. Lett.
{\bf 78}, 1343 (1997).

\bibitem{Zanchi}  D. Zanchi and H.J. Schulz, Phys. Rev. B {\bf 54}, 9509
(1996); ibid. {\bf 61}, 13609 (2000).

\bibitem{Metzner}  C. J. Halboth and W. Metzner, Phys. Rev. B {\bf 61}, 7364
(2000).

\bibitem{SalmHon}  C. Honerkamp, M. Salmhofer, N. Furukawa, and T.M. Rice,
Phys. Rev. B {\bf 63}, 035109 (2001).

\bibitem{VHIKK}  V. Yu. Irkhin, A. A. Katanin, and M. I. Katsnelson, Phys.
Rev. B {\bf 64}, 165107 (2001).

\bibitem{SalmHon1}  C. Honerkamp and M. Salmhofer, Phys. Rev. B{\bf \ 64},
184516 (2001).

\bibitem{Guinea}  J. V. Alvarez, J. Gonzalez, F. Guinea, and M. A. H.
Vozmediano, J. Phys. Soc. Jpn, {\bf 67}, 1868 (1998); cond-mat/9804153
(unpublished); F. Guinea, Nucl. Phys. B {\bf 642}, 407 (2002).

\bibitem{Kanamori}  J. Kanamori, Progr. Theor. Phys. {\bf 30}, 275 (1963).

\bibitem{Metzner1}  C. J. Halboth and W. Metzner, Phys. Rev. Lett. {\bf 85},
5162 (2000).

\bibitem{SalmHon02}  C. Honerkamp, M. Salmhofer, and T. M. Rice, Eur. Phys.
J. B {\bf 27}, 127 (2002).

\bibitem{Shankar}  R. Shankar, Rev. Mod. Phys. {\bf 66}, 129 (1994).

\bibitem{Led}  P. Lederer, G. Montambaux, and D. Poilblanc, J. Phys. (Paris)
{\bf 48}, 1613 (1987).

\bibitem{FurRice}  N. Furukawa, T. M. Rice, and M. Salmhofer, Phys. Rev.
Lett. {\bf 81}, 3195 (1998).

\bibitem{UVJ}  A. P. Kampf and A. A. Katanin, Phys. Rev. B {\bf 67}, 125104
(2003).

\bibitem{Wilson}  K. G. Wilson and J. Kogut, Phys. Rep. {\bf 12}, 77 (1974).

\bibitem{Binz}  B. Binz, D. Baeriswyl, and B. Doucot, Eur. Phys. J. B {\bf 25%
}, 69 (2002).

\bibitem{CHN}  S. Chakravarty, B. I. Halperin, and D. R. Nelson, Phys. Rev.
B {\bf 39}, 2344 (1989).

\bibitem{Kopietz}  P. Kopietz and S. Chakravarty Phys. Rev. B {\bf 40}, 4858
(1989)

\bibitem{Vilk}  A. M. Dare, Y. M. Vilk, and A. M. S. Tremblay, Phys. Rev. B
{\bf 53}, 14236 (1996); Y. M. Vilk and A. M. S. Tremblay, J. Phys. I
(France) {\bf 7}, 1309 (1997).

\bibitem{BKT}  V. L. Berezinskii, ZhETF {\bf 59}, 907 (1970) [Sov. Phys.
JETP {\bf 32, }493 (1970)]; L. M. Kosterlitz and D. J. Thouless, J. Phys. C
{\bf 6}, 1181 (1973).

\bibitem{Kogut}  J. B. Kogut, Rev. Mod. Phys. {\bf 51}, 659 (1979).

\bibitem{Note}  We determine the leading order parameter by the highest
value of the crossover temperature $T_m^{*}$ found from the extrapolation of
the inverse susceptibility, as described in Sect. IIB.

\bibitem{Dongen}  P. G. J. van Dongen, Phys. Rev. B {\bf 54}, 1584 (1996).

\bibitem{PS}  E. Dagotto. J. Riera, Y. C. Chen, A. Moreo, A. Nazarenko, F.
Alcaraz, and F. Ortolani, Phys. Rev. B {\bf 49}, 3548 (1994); F. Guinea, G.
Gomez-Santos, and D. P. Arovas, Phys. Rev. B {\bf 62}, 391 (2000).

\bibitem{MFHalffil}  H. Kondo and T. Moriya, J. Phys. Soc. Jpn. {\bf 65},
2559 (1996).

\bibitem{MFHalffil1}  D. Duffy and A. Moreo, Phys. Rev. B {\bf 55}, R676
(1997).

\bibitem{MFHalffil2}  W. Hofstetter and D. Vollhardt, Ann. Physik {\bf 7},
48 (1998).

\bibitem{PIRG}  T. Kashima and M. Imada, J. Phys. Soc. Jpn. {\bf 70}, 3052
(2001).
\end{references}
\end{document}